\documentclass[12pt]{article}
\usepackage{amsmath,amssymb}
\usepackage{array}
\usepackage{bm}
\usepackage{booktabs}
\usepackage{caption}
\usepackage[backend=biber,style=nature]{biblatex}
\usepackage{xcolor, graphicx}
\usepackage{enumerate}
\usepackage{float,graphicx}
\usepackage{geometry}
\usepackage[utf8]{inputenc}
\usepackage[running]{lineno}
\usepackage{multirow}
\usepackage{mwe}
\usepackage{setspace}
\usepackage{subcaption}
\usepackage{hyperref}
\usepackage[most]{tcolorbox}
\usepackage{lmodern} 
\usepackage[T1]{fontenc}

\usepackage{setspace}

\usepackage[skip=1cm]{parskip}

\newcommand{\Cbb}{\mathbb{C}}
\newcommand{\Ebb}{\mathbb{E}}
\newcommand{\bG}{\mathbf{G}}
\newcommand{\bX}{\mathbf{X}}
\newcommand{\cov}{\text{Cov}}
\newcommand{\diag}{\text{diag}}
\newcommand{\miss}{\text{miss}}

\newcommand{\obs}{\text{obs}}
\newcommand{\Pbb}{\mathbb{P}}

\newcommand{\tr}{\text{tr}}
\newcommand{\Vbb}{\mathbb{V}}
\def\bcdot{\boldsymbol{\cdot} }

\title{Exact Inference for Random Effects Meta-Analyses \\ with Small, Sparse Data}

\author{Jessica Gronsbell$^{1}$, Zachary R. McCaw$^{2}$, Timothy Regis$^{1}$, Lu Tian$^{3}$ \\
\small
Correspondence: \href{mailto:j.gronsbell@utoronto.ca}{j.gronsbell@utoronto.ca}.
}
\date{\small University of Toronto$^{1}$, Harvard School of Public Health$^{2}$, Stanford University$^{3}$ }

\begin{document}

%
%
%
%
%
\def\bzero{{\bf 0}}
\def\bone{{\bf 1}}
%
%
%
%
\def\ba{{\mbox{\boldmath$a$}}}
\def\bb{{\bf b}}
\def\bc{{\bf c}}
\def\bd{{\bf d}}
\def\be{{\bf e}}
\def\bdf{{\bf f}}
\def\bg{{\mbox{\boldmath$g$}}}
\def\bh{{\bf h}}
\def\bi{{\bf i}}
\def\bj{{\bf j}}
\def\bk{{\bf k}}
\def\bl{{\bf l}}
\def\bm{{\bf m}}
\def\bn{{\bf n}}
\def\bo{{\bf o}}
\def\bp{{\bf p}}
\def\bq{{\bf q}}
\def\br{{\bf r}}
\def\bs{{\bf s}}
\def\bt{{\bf t}}
\def\bu{{\bf u}}
\def\bv{{\bf v}}
\def\bw{{\bf w}}
\def\bx{{\bf x}}
\def\by{{\bf y}}
\def\bz{{\bf z}}
\def\bA{{\bf A}}
\def\bB{{\bf B}}
\def\bC{{\bf C}}
\def\bD{{\bf D}}
\def\bE{{\bf E}}
\def\bF{{\bf F}}
\def\bG{{\bf G}}
\def\bH{{\bf H}}
\def\bI{{\bf I}}
\def\bJ{{\bf J}}
\def\bK{{\bf K}}
\def\bL{{\bf L}}
\def\bM{{\bf M}}
\def\bN{{\bf N}}
\def\bO{{\bf O}}
\def\bP{{\bf P}}
\def\bQ{{\bf Q}}
\def\bR{{\bf R}}
\def\bS{{\bf S}}
\def\bT{{\bf T}}
\def\bU{{\bf U}}
\def\bV{{\bf V}}
\def\bW{{\bf W}}
\def\bX{{\bf X}}
\def\bY{{\bf Y}}
\def\bZ{{\bf Z}}
\def\smbZ{\scriptstyle{\bf Z}}
\def\smM{\scriptstyle{M}}
\def\smN{\scriptstyle{N}}
\def\smbT{\scriptstyle{\bf T}}
%
%
%
%
\def\thick#1{\hbox{\rlap{$#1$}\kern0.25pt\rlap{$#1$}\kern0.25pt$#1$}}
\def\balpha{\boldsymbol{\alpha}}
\def\bbeta{\boldsymbol{\beta}}
\def\bgamma{\boldsymbol{\gamma}}
\def\bdelta{\boldsymbol{\delta}}
\def\bepsilon{\boldsymbol{\epsilon}}
\def\bvarepsilon{\boldsymbol{\varepsilon}}
\def\bzeta{\boldsymbol{\zeta}}
\def\bdeta{\boldsymbol{\eta}}
\def\btheta{\boldsymbol{\theta}}
\def\biota{\boldsymbol{\iota}}
\def\bkappa{\boldsymbol{\kappa}}
\def\blambda{\boldsymbol{\lambda}}
\def\bmu{\boldsymbol{\mu}}
\def\bnu{\boldsymbol{\nu}}
\def\bxi{\boldsymbol{\xi}}
\def\bomicron{\boldsymbol{\omicron}}
\def\bpi{\boldsymbol{\pi}}
\def\brho{\boldsymbol{\rho}}
\def\bsigma{\boldsymbol{\sigma}}
\def\btau{\boldsymbol{\tau}}
\def\bupsilon{\boldsymbol{\upsilon}}
\def\bphi{\boldsymbol{\phi}}
\def\bchi{\boldsymbol{\chi}}
\def\bpsi{\boldsymbol{\psi}}
\def\bomega{\boldsymbol{\omega}}
\def\bAlpha{\boldsymbol{\Alpha}}
\def\bBeta{\boldsymbol{\Beta}}
\def\bGamma{\boldsymbol{\Gamma}}
\def\bDelta{\boldsymbol{\Delta}}
\def\bEpsilon{\boldsymbol{\Epsilon}}
\def\bZeta{\boldsymbol{\Zeta}}
\def\bEta{\boldsymbol{\Eta}}
\def\bTheta{\boldsymbol{\Theta}}
\def\bIota{\boldsymbol{\Iota}}
\def\bKappa{\boldsymbol{\Kappa}}
\def\bLambda{{\boldsymbol{\Lambda}}}
\def\bMu{\boldsymbol{\Mu}}
\def\bNu{\boldsymbol{\Nu}}
\def\bXi{\boldsymbol{\Xi}}
\def\bOmicron{\boldsymbol{\Omicron}}
\def\bPi{\boldsymbol{\Pi}}
\def\bRho{\boldsymbol{\Rho}}
\def\bSigma{\boldsymbol{\Sigma}}
\def\bTau{\boldsymbol{\Tau}}
\def\bUpsilon{\boldsymbol{\Upsilon}}
\def\bPhi{\boldsymbol{\Phi}}
\def\bChi{\boldsymbol{\Chi}}
\def\bPsi{\boldsymbol{\Psi}}
\def\bOmega{\boldsymbol{\Omega}}
%
%
%
\def\smalpha{{{\scriptstyle{\alpha}}}}
\def\smbeta{{{\scriptstyle{\beta}}}}
\def\smgamma{{{\scriptstyle{\gamma}}}}
\def\smdelta{{{\scriptstyle{\delta}}}}
\def\smepsilon{{{\scriptstyle{\epsilon}}}}
\def\smvarepsilon{{{\scriptstyle{\varepsilon}}}}
\def\smzeta{{{\scriptstyle{\zeta}}}}
\def\smdeta{{{\scriptstyle{\eta}}}}
\def\smtheta{{{\scriptstyle{\theta}}}}
\def\smiota{{{\scriptstyle{\iota}}}}
\def\smkappa{{{\scriptstyle{\kappa}}}}
\def\smlambda{{{\scriptstyle{\lambda}}}}
\def\smmu{{{\scriptstyle{\mu}}}}
\def\smnu{{{\scriptstyle{\nu}}}}
\def\smxi{{{\scriptstyle{\xi}}}}
\def\smomicron{{{\scriptstyle{\omicron}}}}
\def\smpi{{{\scriptstyle{\pi}}}}
\def\smrho{{{\scriptstyle{\rho}}}}
\def\smsigma{{{\scriptstyle{\sigma}}}}
\def\smtau{{{\scriptstyle{\tau}}}}
\def\smupsilon{{{\scriptstyle{\upsilon}}}}
\def\smphi{{{\scriptstyle{\phi}}}}
\def\smchi{{{\scriptstyle{\chi}}}}
\def\smpsi{{{\scriptstyle{\psi}}}}
\def\smomega{{{\scriptstyle{\omega}}}}
\def\smAlpha{{{\scriptstyle{\Alpha}}}}
\def\smBeta{{{\scriptstyle{\Beta}}}}
\def\smGamma{{{\scriptstyle{\Gamma}}}}
\def\smDelta{{{\scriptstyle{\Delta}}}}
\def\smEpsilon{{{\scriptstyle{\Epsilon}}}}
\def\smZeta{{{\scriptstyle{\Zeta}}}}
\def\smEta{{{\scriptstyle{\Eta}}}}
\def\smTheta{{{\scriptstyle{\Theta}}}}
\def\smIota{{{\scriptstyle{\Iota}}}}
\def\smKappa{{{\scriptstyle{\Kappa}}}}
\def\smLambda{{{\scriptstyle{\Lambda}}}}
\def\smMu{{{\scriptstyle{\Mu}}}}
\def\smNu{{{\scriptstyle{\Nu}}}}
\def\smXi{{{\scriptstyle{\Xi}}}}
\def\smOmicron{{{\scriptstyle{\Omicron}}}}
\def\smPi{{{\scriptstyle{\Pi}}}}
\def\smRho{{{\scriptstyle{\Rho}}}}
\def\smSigma{{{\scriptstyle{\Sigma}}}}
\def\smTau{{{\scriptstyle{\Tau}}}}
\def\smUpsilon{{{\scriptstyle{\Upsilon}}}}
\def\smPhi{{{\scriptstyle{\Phi}}}}
\def\smChi{{{\scriptstyle{\Chi}}}}
\def\smPsi{{{\scriptstyle{\Psi}}}}
\def\smOmega{{{\scriptstyle{\Omega}}}}
%
%

%
\def\smbalpha{\boldsymbol{{\scriptstyle{\alpha}}}}
\def\smbbeta{\boldsymbol{{\scriptstyle{\beta}}}}
\def\smbgamma{\boldsymbol{{\scriptstyle{\gamma}}}}
\def\smbdelta{\boldsymbol{{\scriptstyle{\delta}}}}
\def\smbepsilon{\boldsymbol{{\scriptstyle{\epsilon}}}}
\def\smbvarepsilon{\boldsymbol{{\scriptstyle{\varepsilon}}}}
\def\smbzeta{\boldsymbol{{\scriptstyle{\zeta}}}}
\def\smbdeta{\boldsymbol{{\scriptstyle{\eta}}}}
\def\smbtheta{\boldsymbol{{\scriptstyle{\theta}}}}
\def\smbiota{\boldsymbol{{\scriptstyle{\iota}}}}
\def\smbkappa{\boldsymbol{{\scriptstyle{\kappa}}}}
\def\smblambda{\boldsymbol{{\scriptstyle{\lambda}}}}
\def\smbmu{\boldsymbol{{\scriptstyle{\mu}}}}
\def\smbnu{\boldsymbol{{\scriptstyle{\nu}}}}
\def\smbxi{\boldsymbol{{\scriptstyle{\xi}}}}
\def\smbomicron{\boldsymbol{{\scriptstyle{\omicron}}}}
\def\smbpi{\boldsymbol{{\scriptstyle{\pi}}}}
\def\smbrho{\boldsymbol{{\scriptstyle{\rho}}}}
\def\smbsigma{\boldsymbol{{\scriptstyle{\sigma}}}}
\def\smbtau{\boldsymbol{{\scriptstyle{\tau}}}}
\def\smbupsilon{\boldsymbol{{\scriptstyle{\upsilon}}}}
\def\smbphi{\boldsymbol{{\scriptstyle{\phi}}}}
\def\smbchi{\boldsymbol{{\scriptstyle{\chi}}}}
\def\smbpsi{\boldsymbol{{\scriptstyle{\psi}}}}
\def\smbomega{\boldsymbol{{\scriptstyle{\omega}}}}
\def\smbAlpha{\boldsymbol{{\scriptstyle{\Alpha}}}}
\def\smbBeta{\boldsymbol{{\scriptstyle{\Beta}}}}
\def\smbGamma{\boldsymbol{{\scriptstyle{\Gamma}}}}
\def\smbDelta{\boldsymbol{{\scriptstyle{\Delta}}}}
\def\smbEpsilon{\boldsymbol{{\scriptstyle{\Epsilon}}}}
\def\smbZeta{\boldsymbol{{\scriptstyle{\Zeta}}}}
\def\smbEta{\boldsymbol{{\scriptstyle{\Eta}}}}
\def\smbTheta{\boldsymbol{{\scriptstyle{\Theta}}}}
\def\smbIota{\boldsymbol{{\scriptstyle{\Iota}}}}
\def\smbKappa{\boldsymbol{{\scriptstyle{\Kappa}}}}
\def\smbLambda{\boldsymbol{{\scriptstyle{\Lambda}}}}
\def\smbMu{\boldsymbol{{\scriptstyle{\Mu}}}}
\def\smbNu{\boldsymbol{{\scriptstyle{\Nu}}}}
\def\smbXi{\boldsymbol{{\scriptstyle{\Xi}}}}
\def\smbOmicron{\boldsymbol{{\scriptstyle{\Omicron}}}}
\def\smbPi{\boldsymbol{{\scriptstyle{\Pi}}}}
\def\smbRho{\boldsymbol{{\scriptstyle{\Rho}}}}
\def\smbSigma{\boldsymbol{{\scriptstyle{\Sigma}}}}
\def\smbTau{\boldsymbol{{\scriptstyle{\Tau}}}}
\def\smbUpsilon{\boldsymbol{{\scriptstyle{\Upsilon}}}}
\def\smbPhi{\boldsymbol{{\scriptstyle{\Phi}}}}
\def\smbChi{\boldsymbol{{\scriptstyle{\Chi}}}}
\def\smbPsi{\boldsymbol{{\scriptstyle{\Psi}}}}
\def\smbOmega{\boldsymbol{{\scriptstyle{\Omega}}}}
%
%
%
%
\def\ahat{{\widehat a}}
\def\bhat{{\widehat b}}
\def\chat{{\widehat c}}
\def\dhat{{\widehat d}}
\def\ehat{{\widehat e}}
\def\fhat{{\widehat f}}
\def\ghat{{\widehat g}}
\def\hhat{{\widehat h}}
\def\ihat{{\widehat i}}
\def\jhat{{\widehat j}}
\def\khat{{\widehat k}}
\def\lhat{{\widehat l}}
\def\mhat{{\widehat m}}
\def\nhat{{\widehat n}}
\def\ohat{{\widehat o}}
\def\phat{{\widehat p}}
\def\qhat{{\widehat q}}
\def\rhat{{\widehat r}}
\def\shat{{\widehat s}}
\def\that{{\widehat t}}
\def\uhat{{\widehat u}}
\def\vhat{{\widehat v}}
\def\what{{\widehat w}}
\def\xhat{{\widehat x}}
\def\yhat{{\widehat y}}
\def\zhat{{\widehat z}}
\def\Ahat{{\widehat A}}
\def\Bhat{{\widehat B}}
\def\Chat{{\widehat C}}
\def\Dhat{{\widehat D}}
\def\Ehat{{\widehat E}}
\def\Fhat{{\widehat F}}
\def\Ghat{{\widehat G}}
\def\Hhat{{\widehat H}}
\def\Ihat{{\widehat I}}
\def\Jhat{{\widehat J}}
\def\Khat{{\widehat K}}
\def\Lhat{{\widehat L}}
\def\Mhat{{\widehat M}}
\def\Nhat{{\widehat N}}
\def\Ohat{{\widehat O}}
\def\Phat{{\widehat P}}
\def\Qhat{{\widehat Q}}
\def\Rhat{{\widehat R}}
\def\Shat{{\widehat S}}
\def\That{{\widehat T}}
\def\Uhat{{\widehat U}}
\def\Vhat{{\widehat V}}
\def\What{{\widehat W}}
\def\Xhat{{\widehat X}}
\def\Yhat{{\widehat Y}}
\def\Zhat{{\widehat Z}}
%
%
%
\def\atilde{{\widetilde a}}
\def\btilde{{\widetilde b}}
\def\ctilde{{\widetilde c}}
\def\dtilde{{\widetilde d}}
\def\etilde{{\widetilde e}}
\def\ftilde{{\widetilde f}}
\def\gtilde{{\widetilde g}}
\def\htilde{{\widetilde h}}
\def\itilde{{\widetilde i}}
\def\jtilde{{\widetilde j}}
\def\ktilde{{\widetilde k}}
\def\ltilde{{\widetilde l}}
\def\mtilde{{\widetilde m}}
\def\ntilde{{\widetilde n}}
\def\otilde{{\widetilde o}}
\def\ptilde{{\widetilde p}}
\def\qtilde{{\widetilde q}}
\def\rtilde{{\widetilde r}}
\def\stilde{{\widetilde s}}
\def\ttilde{{\widetilde t}}
\def\utilde{{\widetilde u}}
\def\vtilde{{\widetilde v}}
\def\wtilde{{\widetilde w}}
\def\xtilde{{\widetilde x}}
\def\ytilde{{\widetilde y}}
\def\ztilde{{\widetilde z}}
\def\Atilde{{\widetilde A}}
\def\Btilde{{\widetilde B}}
\def\Ctilde{{\widetilde C}}
\def\Dtilde{{\widetilde D}}
\def\Etilde{{\widetilde E}}
\def\Ftilde{{\widetilde F}}
\def\Gtilde{{\widetilde G}}
\def\Htilde{{\widetilde H}}
\def\Itilde{{\widetilde I}}
\def\Jtilde{{\widetilde J}}
\def\Ktilde{{\widetilde K}}
\def\Ltilde{{\widetilde L}}
\def\Mtilde{{\widetilde M}}
\def\Ntilde{{\widetilde N}}
\def\Otilde{{\widetilde O}}
\def\Ptilde{{\widetilde P}}
\def\Qtilde{{\widetilde Q}}
\def\Rtilde{{\widetilde R}}
\def\Stilde{{\widetilde S}}
\def\Ttilde{{\widetilde T}}
\def\Utilde{{\widetilde U}}
\def\Vtilde{{\widetilde V}}
\def\Wtilde{{\widetilde W}}
\def\Xtilde{{\widetilde X}}
\def\Ytilde{{\widetilde Y}}
\def\Ztilde{{\widetilde Z}}
%
%
%
%
\def\bahat{{\widehat \ba}}
\def\bbhat{{\widehat \bb}}
\def\bchat{{\widehat \bc}}
\def\bdhat{{\widehat \bd}}
\def\behat{{\widehat \be}}
\def\bfhat{{\widehat \bf}}
\def\bghat{{\widehat \bg}}
\def\bhhat{{\widehat \bh}}
\def\bihat{{\widehat \bi}}
\def\bjhat{{\widehat \bj}}
\def\bkhat{{\widehat \bk}}
\def\blhat{{\widehat \bl}}
\def\bmhat{{\widehat \bm}}
\def\bnhat{{\widehat \bn}}
\def\bohat{{\widehat \bo}}
\def\bphat{{\widehat \bp}}
\def\bqhat{{\widehat \bq}}
\def\brhat{{\widehat \br}}
\def\bshat{{\widehat \bs}}
\def\bthat{{\widehat \bt}}
\def\buhat{{\widehat \bu}}
\def\bvhat{{\widehat \bv}}
\def\bwhat{{\widehat \bw}}
\def\bxhat{{\widehat \bx}}
\def\byhat{{\widehat \by}}
\def\bzhat{{\widehat \bz}}
\def\bAhat{{\widehat \bA}}
\def\bBhat{{\widehat \bB}}
\def\bChat{{\widehat \bC}}
\def\bDhat{{\widehat \bD}}
\def\bEhat{{\widehat \bE}}
\def\bFhat{{\widehat \bF}}
\def\bGhat{{\widehat \bG}}
\def\bHhat{{\widehat \bH}}
\def\bIhat{{\widehat \bI}}
\def\bJhat{{\widehat \bJ}}
\def\bKhat{{\widehat \bK}}
\def\bLhat{{\widehat \bL}}
\def\bMhat{{\widehat \bM}}
\def\bNhat{{\widehat \bN}}
\def\bOhat{{\widehat \bO}}
\def\bPhat{{\widehat \bP}}
\def\bQhat{{\widehat \bQ}}
\def\bRhat{{\widehat \bR}}
\def\bShat{{\widehat \bS}}
\def\bThat{{\widehat \bT}}
\def\bUhat{{\widehat \bU}}
\def\bVhat{{\widehat \bV}}
\def\bWhat{{\widehat \bW}}
\def\bXhat{{\widehat \bX}}
\def\bYhat{{\widehat \bY}}
\def\bZhat{{\widehat \bZ}}
%
%
%
%
%
\def\batilde{{\widetilde \ba}}
\def\bbtilde{{\widetilde \bb}}
\def\bctilde{{\widetilde \bc}}
\def\bdtilde{{\widetilde \bd}}
\def\betilde{{\widetilde \be}}
\def\bftilde{{\widetilde \bf}}
\def\bgtilde{{\widetilde \bg}}
\def\bhtilde{{\widetilde \bh}}
\def\bitilde{{\widetilde \bi}}
\def\bjtilde{{\widetilde \bj}}
\def\bktilde{{\widetilde \bk}}
\def\bltilde{{\widetilde \bl}}
\def\bmtilde{{\widetilde \bm}}
\def\bntilde{{\widetilde \bn}}
\def\botilde{{\widetilde \bo}}
\def\bptilde{{\widetilde \bp}}
\def\bqtilde{{\widetilde \bq}}
\def\brtilde{{\widetilde \br}}
\def\bstilde{{\widetilde \bs}}
\def\bttilde{{\widetilde \bt}}
\def\butilde{{\widetilde \bu}}
\def\bvtilde{{\widetilde \bv}}
\def\bwtilde{{\widetilde \bw}}
\def\bxtilde{{\widetilde \bx}}
\def\bytilde{{\widetilde \by}}
\def\bztilde{{\widetilde \bz}}
\def\bAtilde{{\widetilde \bA}}
\def\bBtilde{{\widetilde \bB}}
\def\bCtilde{{\widetilde \bC}}
\def\bDtilde{{\widetilde \bD}}
\def\bEtilde{{\widetilde \bE}}
\def\bFtilde{{\widetilde \bF}}
\def\bGtilde{{\widetilde \bG}}
\def\bHtilde{{\widetilde \bH}}
\def\bItilde{{\widetilde \bI}}
\def\bJtilde{{\widetilde \bJ}}
\def\bKtilde{{\widetilde \bK}}
\def\bLtilde{{\widetilde \bL}}
\def\bMtilde{{\widetilde \bM}}
\def\bNtilde{{\widetilde \bN}}
\def\bOtilde{{\widetilde \bO}}
\def\bPtilde{{\widetilde \bP}}
\def\bQtilde{{\widetilde \bQ}}
\def\bRtilde{{\widetilde \bR}}
\def\bStilde{{\widetilde \bS}}
\def\bTtilde{{\widetilde \bT}}
\def\bUtilde{{\widetilde \bU}}
\def\bVtilde{{\widetilde \bV}}
\def\bWtilde{{\widetilde \bW}}
\def\bXtilde{{\widetilde \bX}}
\def\bYtilde{{\widetilde \bY}}
\def\bZtilde{{\widetilde \bZ}}
%
%
%
%
%
%
\def\alphahat{{\widehat\alpha}}
\def\betahat{{\widehat\beta}}
\def\gammahat{{\widehat\gamma}}
\def\deltahat{{\widehat\delta}}
\def\epsilonhat{{\widehat\epsilon}}
\def\varepsilonhat{{\widehat\varepsilon}}
\def\zetahat{{\widehat\zeta}}
\def\etahat{{\widehat\eta}}
\def\thetahat{{\widehat\theta}}
\def\iotahat{{\widehat\iota}}
\def\kappahat{{\widehat\kappa}}
\def\lambdahat{{\widehat\lambda}}
\def\muhat{{\widehat\mu}}
\def\nuhat{{\widehat\nu}}
\def\xihat{{\widehat\xi}}
\def\omicronhat{{\widehat\omicron}}
\def\pihat{{\widehat\pi}}
\def\rhohat{{\widehat\rho}}
\def\sigmahat{{\widehat\sigma}}
\def\tauhat{{\widehat\tau}}
\def\upsilonhat{{\widehat\upsilon}}
\def\phihat{{\widehat\phi}}
\def\chihat{{\widehat\chi}}
\def\psihat{{\widehat\psi}}
\def\omegahat{{\widehat\omega}}
\def\Alphahat{{\widehat\Alpha}}
\def\Betahat{{\widehat\Beta}}
\def\Gammahat{{\widehat\Gamma}}
\def\Deltahat{{\widehat\Delta}}
\def\Epsilonhat{{\widehat\Epsilon}}
\def\Zetahat{{\widehat\Zeta}}
\def\Etahat{{\widehat\Eta}}
\def\Thetahat{{\widehat\Theta}}
\def\Iotahat{{\widehat\Iota}}
\def\Kappahat{{\widehat\Kappa}}
\def\Lambdahat{{\widehat\Lambda}}
\def\Muhat{{\widehat\Mu}}
\def\Nuhat{{\widehat\Nu}}
\def\Xihat{{\widehat\Xi}}
\def\Omicronhat{{\widehat\Omicron}}
\def\Pihat{{\widehat\Pi}}
\def\Rhohat{{\widehat\Rho}}
\def\Sigmahat{{\widehat\Sigma}}
\def\Tauhat{{\widehat\Tau}}
\def\Upsilonhat{{\widehat\Upsilon}}
\def\Phihat{{\widehat\Phi}}
\def\Chihat{{\widehat\Chi}}
\def\Psihat{{\widehat\Psi}}
\def\Omegahat{{\widehat\Omega}}
%
%
%
%
%
\def\alphatilde{{\widetilde\alpha}}
\def\betatilde{{\widetilde\beta}}
\def\gammatilde{{\widetilde\gamma}}
\def\deltatilde{{\widetilde\delta}}
\def\epsilontilde{{\widetilde\epsilon}}
\def\varepsilontilde{{\widetilde\varepsilon}}
\def\zetatilde{{\widetilde\zeta}}
\def\etatilde{{\widetilde\eta}}
\def\thetatilde{{\widetilde\theta}}
\def\iotatilde{{\widetilde\iota}}
\def\kappatilde{{\widetilde\kappa}}
\def\lambdatilde{{\widetilde\lambda}}
\def\mutilde{{\widetilde\mu}}
\def\nutilde{{\widetilde\nu}}
\def\xitilde{{\widetilde\xi}}
\def\omicrontilde{{\widetilde\omicron}}
\def\pitilde{{\widetilde\pi}}
\def\rhotilde{{\widetilde\rho}}
\def\sigmatilde{{\widetilde\sigma}}
\def\tautilde{{\widetilde\tau}}
\def\upsilontilde{{\widetilde\upsilon}}
\def\phitilde{{\widetilde\phi}}
\def\chitilde{{\widetilde\chi}}
\def\psitilde{{\widetilde\psi}}
\def\omegatilde{{\widetilde\omega}}
\def\Alphatilde{{\widetilde\Alpha}}
\def\Betatilde{{\widetilde\Beta}}
\def\Gammatilde{{\widetilde\Gamma}}
\def\Deltatilde{{\widetilde\Delta}}
\def\Epsilontilde{{\widetilde\Epsilon}}
\def\Zetatilde{{\widetilde\Zeta}}
\def\Etatilde{{\widetilde\Eta}}
\def\Thetatilde{{\widetilde\Theta}}
\def\Iotatilde{{\widetilde\Iota}}
\def\Kappatilde{{\widetilde\Kappa}}
\def\Lambdatilde{{\widetilde\Lambda}}
\def\Mutilde{{\widetilde\Mu}}
\def\Nutilde{{\widetilde\Nu}}
\def\Xitilde{{\widetilde\Xi}}
\def\Omicrontilde{{\widetilde\Omicron}}
\def\Pitilde{{\widetilde\Pi}}
\def\Rhotilde{{\widetilde\Rho}}
\def\Sigmatilde{{\widetilde\Sigma}}
\def\Tautilde{{\widetilde\Tau}}
\def\Upsilontilde{{\widetilde\Upsilon}}
\def\Phitilde{{\widetilde\Phi}}
\def\Chitilde{{\widetilde\Chi}}
\def\Psitilde{{\widetilde\Psi}}
\def\Omegatilde{{\widetilde\Omega}}
%
%
%
%
%
%
\def\balphahat{{\widehat\balpha}}
\def\bbetahat{{\widehat\bbeta}}
\def\bgammahat{{\widehat\bgamma}}
\def\bdeltahat{{\widehat\bdelta}}
\def\bepsilonhat{{\widehat\bepsilon}}
\def\bzetahat{{\widehat\bzeta}}
\def\bdetahat{{\widehat\bdeta}}
\def\bthetahat{{\widehat\btheta}}
\def\biotahat{{\widehat\biota}}
\def\bkappahat{{\widehat\bkappa}}
\def\blambdahat{{\widehat\blambda}}
\def\bmuhat{{\widehat\bmu}}
\def\bnuhat{{\widehat\bnu}}
\def\bxihat{{\widehat\bxi}}
\def\bomicronhat{{\widehat\bomicron}}
\def\bpihat{{\widehat\bpi}}
\def\brhohat{{\widehat\brho}}
\def\bsigmahat{{\widehat\bsigma}}
\def\btauhat{{\widehat\btau}}
\def\bupsilonhat{{\widehat\bupsilon}}
\def\bphihat{{\widehat\bphi}}
\def\bchihat{{\widehat\bchi}}
\def\bpsihat{{\widehat\bpsi}}
\def\bomegahat{{\widehat\bomega}}
\def\bAlphahat{{\widehat\bAlpha}}
\def\bBetahat{{\widehat\bBeta}}
\def\bGammahat{{\widehat\bGamma}}
\def\bDeltahat{{\widehat\bDelta}}
\def\bEpsilonhat{{\widehat\bEpsilon}}
\def\bZetahat{{\widehat\bZeta}}
\def\bEtahat{{\widehat\bEta}}
\def\bThetahat{{\widehat\bTheta}}
\def\bIotahat{{\widehat\bIota}}
\def\bKappahat{{\widehat\bKappa}}
\def\bLambdahat{{\widehat\bLambda}}
\def\bMuhat{{\widehat\bMu}}
\def\bNuhat{{\widehat\bNu}}
\def\bXihat{{\widehat\bXi}}
\def\bOmicronhat{{\widehat\bOmicron}}
\def\bPihat{{\widehat\bPi}}
\def\bRhohat{{\widehat\bRho}}
\def\bSigmahat{{\widehat\bSigma}}
\def\bTauhat{{\widehat\bTau}}
\def\bUpsilonhat{{\widehat\bUpsilon}}
\def\bPhihat{{\widehat\bPhi}}
\def\bChihat{{\widehat\bChi}}
\def\bPsihat{{\widehat\bPsi}}
\def\bOmegahat{{\widehat\bOmega}}%
\def\balphahattrans{{\balphahat^{_{\transpose}}}}
\def\bbetahattrans{{\bbetahat^{_{\transpose}}}}
\def\bgammahattrans{{\bgammahat^{_{\transpose}}}}
\def\bdeltahattrans{{\bdeltahat^{_{\transpose}}}}
\def\bepsilonhattrans{{\bepsilonhat^{_{\transpose}}}}
\def\bzetahattrans{{\bzetahat^{_{\transpose}}}}
\def\bdetahattrans{{\bdetahat^{_{\transpose}}}}
\def\bthetahattrans{{\bthetahat^{_{\transpose}}}}
\def\biotahattrans{{\biotahat^{_{\transpose}}}}
\def\bkappahattrans{{\bkappahat^{_{\transpose}}}}
\def\blambdahattrans{{\blambdahat^{_{\transpose}}}}
\def\bmuhattrans{{\bmuhat^{_{\transpose}}}}
\def\bnuhattrans{{\bnuhat^{_{\transpose}}}}
\def\bxihattrans{{\bxihat^{_{\transpose}}}}
\def\bomicronhattrans{{\bomicronhat^{_{\transpose}}}}
\def\bpihattrans{{\bpihat^{_{\transpose}}}}
\def\brhohattrans{{\brhohat^{_{\transpose}}}}
\def\bsigmahattrans{{\bsigmahat^{_{\transpose}}}}
\def\btauhattrans{{\btauhat^{_{\transpose}}}}
\def\bupsilonhattrans{{\bupsilonhat^{_{\transpose}}}}
\def\bphihattrans{{\bphihat^{_{\transpose}}}}
\def\bchihattrans{{\bchihat^{_{\transpose}}}}
\def\bpsihattrans{{\bpsihat^{_{\transpose}}}}
\def\bomegahattrans{{\bomegahat^{_{\transpose}}}}
\def\bAlphahattrans{{\bAlphahat^{_{\transpose}}}}
\def\bBetahattrans{{\bBetahat^{_{\transpose}}}}
\def\bGammahattrans{{\bGammahat^{_{\transpose}}}}
\def\bDeltahattrans{{\bDeltahat^{_{\transpose}}}}
\def\bEpsilonhattrans{{\bEpsilonhat^{_{\transpose}}}}
\def\bZetahattrans{{\bZetahat^{_{\transpose}}}}
\def\bEtahattrans{{\bEtahat^{_{\transpose}}}}
\def\bThetahattrans{{\bThetahat^{_{\transpose}}}}
\def\bIotahattrans{{\bIotahat^{_{\transpose}}}}
\def\bKappahattrans{{\bKappahat^{_{\transpose}}}}
\def\bLambdahattrans{{\bLambdahat^{_{\transpose}}}}
\def\bMuhattrans{{\bMuhat^{_{\transpose}}}}
\def\bNuhattrans{{\bNuhat^{_{\transpose}}}}
\def\bXihattrans{{\bXihat^{_{\transpose}}}}
\def\bOmicronhattrans{{\bOmicronhat^{_{\transpose}}}}
\def\bPihattrans{{\bPihat^{_{\transpose}}}}
\def\bRhohattrans{{\bRhohat^{_{\transpose}}}}
\def\bSigmahattrans{{\bSigmahat^{_{\transpose}}}}
\def\bTauhattrans{{\bTauhat^{_{\transpose}}}}
\def\bUpsilonhattrans{{\bUpsilonhat^{_{\transpose}}}}
\def\bPhihattrans{{\bPhihat^{_{\transpose}}}}
\def\bChihattrans{{\bChihat^{_{\transpose}}}}
\def\bPsihattrans{{\bPsihat^{_{\transpose}}}}
\def\bOmegahattrans{{\bOmegahat^{_{\transpose}}}}%
%
\def\smbalpha{\widehat{\smbalpha}}
\def\smbbetahat{\widehat{\smbbeta}}
\def\smbgammahat{\widehat{\smbgamma}}
\def\smbdeltahat{\widehat{\smbdelta}}
\def\smbepsilonhat{\widehat{\smbepsilon}}
\def\smbvarepsilonhat{\widehat{\smbvarepsilon}}
\def\smbzetahat{\widehat{\smbzeta}}
\def\smbdetahat{\widehat{\smbeta}}
\def\smbthetahat{\widehat{\smbtheta}}
\def\smbiotahat{\widehat{\smbiota}}
\def\smbkappahat{\widehat{\smbkappa}}
\def\smblambdahat{\widehat{\smblambda}}
\def\smbmuhat{\widehat{\smbmu}}
\def\smbnuhat{\widehat{\smbnu}}
\def\smbxihat{\widehat{\smbxi}}
\def\smbomicronhat{\widehat{\smbomicron}}
\def\smbpihat{\widehat{\smbpi}}
\def\smbrhohat{\widehat{\smbrho}}
\def\smbsigmahat{\widehat{\smbsigma}}
\def\smbtauhat{\widehat{\smbtau}}
\def\smbupsilonhat{\widehat{\smbupsilon}}
\def\smbphihat{\widehat{\smbphi}}
\def\smbchihat{\widehat{\smbchi}}
\def\smbpsihat{\widehat{\smbpsi}}
\def\smbomegahat{\widehat{\smbomega}}
\def\smbAlphahat{\widehat{\smbAlpha}}
\def\smbBetahat{\widehat{\smbBeta}}
\def\smbGammahat{\widehat{\smbGamma}}
\def\smbDeltahat{\widehat{\smbDelta}}
\def\smbEpsilonhat{\widehat{\smbEpsilon}}
\def\smbZetahat{\widehat{\smbZeta}}
\def\smbEtahat{\widehat{\smbEta}}
\def\smbThetahat{\widehat{\smbTheta}}
\def\smbIotahat{\widehat{\smbIota}}
\def\smbKappahat{\widehat{\smbKappa}}
\def\smbLambdahat{\widehat{\smbLambda}}
\def\smbMuhat{\widehat{\smbMu}}
\def\smbNuhat{\widehat{\smbNu}}
\def\smbXihat{\widehat{\smbXi}}
\def\smbOmicronhat{\widehat{\smbOmicron}}
\def\smbPihat{\widehat{\smbPi}}
\def\smbRhohat{\widehat{\smbRho}}
\def\smbSigmahat{\widehat{\smbSigma}}
\def\smbTauhat{\widehat{\smbTau}}
\def\smbUpsilonhat{\widehat{\smbUpsilon}}
\def\smbPhihat{\widehat{\smbPhi}}
\def\smbChihat{\widehat{\smbChi}}
\def\smbPsihat{\widehat{\smbPsi}}
\def\smbOmegahat{\widehat{\smbOmega}}
%
%
%
%
%
\def\balphatilde{{\widetilde\balpha}}
\def\bbetatilde{{\widetilde\bbeta}}
\def\bgammatilde{{\widetilde\bgamma}}
\def\bdeltatilde{{\widetilde\bdelta}}
\def\bepsilontilde{{\widetilde\bepsilon}}
\def\bzetatilde{{\widetilde\bzeta}}
\def\bdetatilde{{\widetilde\bdeta}}
\def\bthetatilde{{\widetilde\btheta}}
\def\biotatilde{{\widetilde\biota}}
\def\bkappatilde{{\widetilde\bkappa}}
\def\blambdatilde{{\widetilde\blambda}}
\def\bmutilde{{\widetilde\bmu}}
\def\bnutilde{{\widetilde\bnu}}
\def\bxitilde{{\widetilde\bxi}}
\def\bomicrontilde{{\widetilde\bomicron}}
\def\bpitilde{{\widetilde\bpi}}
\def\brhotilde{{\widetilde\brho}}
\def\bsigmatilde{{\widetilde\bsigma}}
\def\btautilde{{\widetilde\btau}}
\def\bupsilontilde{{\widetilde\bupsilon}}
\def\bphitilde{{\widetilde\bphi}}
\def\bchitilde{{\widetilde\bchi}}
\def\bpsitilde{{\widetilde\bpsi}}
\def\bomegatilde{{\widetilde\bomega}}
\def\bAlphatilde{{\widetilde\bAlpha}}
\def\bBetatilde{{\widetilde\bBeta}}
\def\bGammatilde{{\widetilde\bGamma}}
\def\bDeltatilde{{\widetilde\bDelta}}
\def\bEpsilontilde{{\widetilde\bEpsilon}}
\def\bZetatilde{{\widetilde\bZeta}}
\def\bEtatilde{{\widetilde\bEta}}
\def\bThetatilde{{\widetilde\bTheta}}
\def\bIotatilde{{\widetilde\bIota}}
\def\bKappatilde{{\widetilde\bKappa}}
\def\bLambdatilde{{\widetilde\bLambda}}
\def\bMutilde{{\widetilde\bMu}}
\def\bNutilde{{\widetilde\bNu}}
\def\bXitilde{{\widetilde\bXi}}
\def\bOmicrontilde{{\widetilde\bOmicron}}
\def\bPitilde{{\widetilde\bPi}}
\def\bRhotilde{{\widetilde\bRho}}
\def\bSigmatilde{{\widetilde\bSigma}}
\def\bTautilde{{\widetilde\bTau}}
\def\bUpsilontilde{{\widetilde\bUpsilon}}
\def\bPhitilde{{\widetilde\bPhi}}
\def\bChitilde{{\widetilde\bChi}}
\def\bPsitilde{{\widetilde\bPsi}}
\def\bOmegatilde{{\widetilde\bOmega}}
%
%
%
%
%
\def\abar{\bar{ a}}
\def\bbar{\bar{ b}}
\def\cbar{\bar{ c}}
\def\dbar{\bar{ d}}
\def\ebar{\bar{ e}}
\def\fbar{\bar{ f}}
\def\gbar{\bar{ g}}
\def\hbar{\bar{ h}}
\def\ibar{\bar{ i}}
\def\jbar{\bar{ j}}
\def\kbar{\bar{ k}}
\def\lbar{\bar{ l}}
\def\mbar{\bar{ m}}
\def\nbar{\bar{ n}}
\def\obar{\bar{ o}}
\def\pbar{\bar{ p}}
\def\qbar{\bar{ q}}
\def\rbar{\bar{ r}}
\def\sbar{\bar{ s}}
\def\tbar{\bar{ t}}
\def\ubar{\bar{ u}}
\def\vbar{\bar{ v}}
\def\wbar{\bar{ w}}
\def\xbar{\bar{ x}}
\def\ybar{\bar{ y}}
\def\zbar{\bar{ z}}
\def\Abar{\bar{ A}}
\def\Bbar{\bar{ B}}
\def\Cbar{\bar{ C}}
\def\Dbar{\bar{ D}}
\def\Ebar{\bar{ E}}
\def\Fbar{\bar{ F}}
\def\Gbar{\bar{ G}}
\def\Hbar{\bar{ H}}
\def\Ibar{\bar{ I}}
\def\Jbar{\bar{ J}}
\def\Kbar{\bar{ K}}
\def\Lbar{\bar{ L}}
\def\Mbar{\bar{ M}}
\def\Nbar{\bar{ N}}
\def\Obar{\bar{ O}}
\def\Pbar{\bar{ P}}
\def\Qbar{\bar{ Q}}
\def\Rbar{\bar{ R}}
\def\Sbar{\bar{ S}}
\def\Tbar{\bar{ T}}
\def\Ubar{\bar{ U}}
\def\Vbar{\bar{ V}}
\def\Wbar{\bar{ W}}
\def\Xbar{\bar{ X}}
\def\Ybar{\bar{ Y}}
\def\Zbar{\bar{ Z}}
%
%
%
%
%
\def\babar{\overline{ \ba}}
\def\bbbar{\overline{ \bb}}
\def\bcbar{\overline{ \bc}}
\def\bdbar{\overline{ \bd}}
\def\bebar{\overline{ \be}}
\def\bfbar{\overline{ \bf}}
\def\bgbar{\overline{ \bg}}
\def\bhbar{\overline{ \bh}}
\def\bibar{\overline{ \bi}}
\def\bjbar{\overline{ \bj}}
\def\bkbar{\overline{ \bk}}
\def\blbar{\overline{ \bl}}
\def\bmbar{\overline{ \bm}}
\def\bnbar{\overline{ \bn}}
\def\bobar{\overline{ \bo}}
\def\bpbar{\overline{ \bp}}
\def\bqbar{\overline{ \bq}}
\def\brbar{\overline{ \br}}
\def\bsbar{\overline{ \bs}}
\def\btbar{\overline{ \bt}}
\def\bubar{\overline{ \bu}}
\def\bvbar{\overline{ \bv}}
\def\bwbar{\overline{ \bw}}
\def\bxbar{\overline{ \bx}}
\def\bybar{\overline{ \by}}
\def\bzbar{\overline{ \bz}}
\def\bAbar{\overline{ \bA}}
\def\bBbar{\overline{ \bB}}
\def\bCbar{\overline{ \bC}}
\def\bDbar{\overline{ \bD}}
\def\bEbar{\overline{ \bE}}
\def\bFbar{\overline{ \bF}}
\def\bGbar{\overline{ \bG}}
\def\bHbar{\overline{ \bH}}
\def\bIbar{\overline{ \bI}}
\def\bJbar{\overline{ \bJ}}
\def\bKbar{\overline{ \bK}}
\def\bLbar{\overline{ \bL}}
\def\bMbar{\overline{ \bM}}
\def\bNbar{\overline{ \bN}}
\def\bObar{\overline{ \bO}}
\def\bPbar{\overline{ \bP}}
\def\bQbar{\overline{ \bQ}}
\def\bRbar{\overline{ \bR}}
\def\bSbar{\overline{ \bS}}
\def\bTbar{\overline{ \bT}}
\def\bUbar{\overline{ \bU}}
\def\bVbar{\overline{ \bV}}
\def\bWbar{\overline{ \bW}}
\def\bXbar{\overline{ \bX}}
\def\bYbar{\overline{ \bY}}
\def\bZbar{\overline{ \bZ}}
%
%

%
%
%
\def\asc{{\cal a}}
\def\bsc{{\cal b}}
\def\csc{{\cal c}}
\def\dsc{{\cal d}}
\def\esc{{\cal e}}
\def\dsc{{\cal f}}
\def\gsc{{\cal g}}
\def\hsc{{\cal h}}
\def\isc{{\cal i}}
\def\jsc{{\cal j}}
\def\ksc{{\cal k}}
\def\lsc{{\cal l}}
\def\msc{{\cal m}}
\def\nsc{{\cal n}}
\def\osc{{\cal o}}
\def\psc{{\cal p}}
\def\qsc{{\cal q}}
\def\rsc{{\cal r}}
\def\ssc{{\cal s}}
\def\tsc{{\cal t}}
\def\usc{{\cal u}}
\def\vsc{{\cal v}}
\def\wsc{{\cal w}}
\def\xsc{{\cal x}}
\def\ysc{{\cal y}}
\def\zsc{{\cal z}}
\def\Asc{{\cal A}}
\def\Bsc{{\cal B}}
\def\Csc{{\cal C}}
\def\Dsc{{\cal D}}
\def\Esc{{\cal E}}
\def\Fsc{{\cal F}}
\def\Gsc{{\cal G}}
\def\Hsc{{\cal H}}
\def\Isc{{\cal I}}
\def\Jsc{{\cal J}}
\def\Ksc{{\cal K}}
\def\Lsc{{\cal L}}
\def\Msc{{\cal M}}
\def\Nsc{{\cal N}}
\def\Osc{{\cal O}}
\def\Psc{{\cal P}}
\def\Qsc{{\cal Q}}
\def\Rsc{{\cal R}}
\def\Ssc{{\cal S}}
\def\Tsc{{\cal T}}
\def\Usc{{\cal U}}
\def\Vsc{{\cal V}}
\def\Wsc{{\cal W}}
\def\Xsc{{\cal X}}
\def\Ysc{{\cal Y}}
\def\Zsc{{\cal Z}}
\def\Aschat{\widehat{{\cal A}}}
\def\Bschat{\widehat{{\cal B}}}
\def\Cschat{\widehat{{\cal C}}}
\def\Dschat{\widehat{{\cal D}}}
\def\Eschat{\widehat{{\cal E}}}
\def\Fschat{\widehat{{\cal F}}}
\def\Gschat{\widehat{{\cal G}}}
\def\Hschat{\widehat{{\cal H}}}
\def\Ischat{\widehat{{\cal I}}}
\def\Jschat{\widehat{{\cal J}}}
\def\Kschat{\widehat{{\cal K}}}
\def\Lschat{\widehat{{\cal L}}}
\def\Mschat{\widehat{{\cal M}}}
\def\Nschat{\widehat{{\cal N}}}
\def\Oschat{\widehat{{\cal O}}}
\def\Pschat{\widehat{{\cal P}}}
\def\Qschat{\widehat{{\cal Q}}}
\def\Rschat{\widehat{{\cal R}}}
\def\Sschat{\widehat{{\cal S}}}
\def\Tschat{\widehat{{\cal T}}}
\def\Uschat{\widehat{{\cal U}}}
\def\Vschat{\widehat{{\cal V}}}
\def\Wschat{\widehat{{\cal W}}}
\def\Xschat{\widehat{{\cal X}}}
\def\Yschat{\widehat{{\cal Y}}}
\def\Zschat{\widehat{{\cal Z}}}
\def\Asctilde{\widetilde{{\cal A}}}
\def\Bsctilde{\widetilde{{\cal B}}}
\def\Csctilde{\widetilde{{\cal C}}}
\def\Dsctilde{\widetilde{{\cal D}}}
\def\Esctilde{\widetilde{{\cal E}}}
\def\Fsctilde{\widetilde{{\cal F}}}
\def\Gsctilde{\widetilde{{\cal G}}}
\def\Hsctilde{\widetilde{{\cal H}}}
\def\Isctilde{\widetilde{{\cal I}}}
\def\Jsctilde{\widetilde{{\cal J}}}
\def\Ksctilde{\widetilde{{\cal K}}}
\def\Lsctilde{\widetilde{{\cal L}}}
\def\Msctilde{\widetilde{{\cal M}}}
\def\Nsctilde{\widetilde{{\cal N}}}
\def\Osctilde{\widetilde{{\cal O}}}
\def\Psctilde{\widetilde{{\cal P}}}
\def\Qsctilde{\widetilde{{\cal Q}}}
\def\Rsctilde{\widetilde{{\cal R}}}
\def\Ssctilde{\widetilde{{\cal S}}}
\def\Tsctilde{\widetilde{{\cal T}}}
\def\Usctilde{\widetilde{{\cal U}}}
\def\Vsctilde{\widetilde{{\cal V}}}
\def\Wsctilde{\widetilde{{\cal W}}}
\def\Xsctilde{\widetilde{{\cal X}}}
\def\Ysctilde{\widetilde{{\cal Y}}}
\def\Zsctilde{\widetilde{{\cal Z}}}
\def\bAsc{\mathbf{\cal A}}
\def\bBsc{\mathbf{\cal B}}
\def\bCsc{\mathbf{\cal C}}
\def\bDsc{\mathbf{\cal D}}
\def\bEsc{\mathbf{\cal E}}
\def\bFsc{\mathbf{\cal F}}
\def\bGsc{\mathbf{\cal G}}
\def\bHsc{\mathbf{\cal H}}
\def\bIsc{\mathbf{\cal I}}
\def\bJsc{\mathbf{\cal J}}
\def\bKsc{\mathbf{\cal K}}
\def\bLsc{\mathbf{\cal L}}
\def\bMsc{\mathbf{\cal M}}
\def\bNsc{\mathbf{\cal N}}
\def\bOsc{\mathbf{\cal O}}
\def\bPsc{\mathbf{\cal P}}
\def\bQsc{\mathbf{\cal Q}}
\def\bRsc{\mathbf{\cal R}}
\def\bSsc{\mathbf{\cal S}}
\def\bTsc{\mathbf{\cal T}}
\def\bUsc{\mathbf{\cal U}}
\def\bVsc{\mathbf{\cal V}}
\def\bWsc{\mathbf{\cal W}}
\def\bXsc{\mathbf{\cal X}}
\def\bYsc{\mathbf{\cal Y}}
\def\bZsc{\mathbf{\cal Z}}
\def\bAschat{\widehat{\mathbf{\cal A}}}
\def\bBschat{\widehat{\mathbf{\cal B}}}
\def\bCschat{\widehat{\mathbf{\cal C}}}
\def\bDschat{\widehat{\mathbf{\cal D}}}
\def\bEschat{\widehat{\mathbf{\cal E}}}
\def\bFschat{\widehat{\mathbf{\cal F}}}
\def\bGschat{\widehat{\mathbf{\cal G}}}
\def\bHschat{\widehat{\mathbf{\cal H}}}
\def\bIschat{\widehat{\mathbf{\cal I}}}
\def\bJschat{\widehat{\mathbf{\cal J}}}
\def\bKschat{\widehat{\mathbf{\cal K}}}
\def\bLschat{\widehat{\mathbf{\cal L}}}
\def\bMschat{\widehat{\mathbf{\cal M}}}
\def\bNschat{\widehat{\mathbf{\cal N}}}
\def\bOschat{\widehat{\mathbf{\cal O}}}
\def\bPschat{\widehat{\mathbf{\cal P}}}
\def\bQschat{\widehat{\mathbf{\cal Q}}}
\def\bRschat{\widehat{\mathbf{\cal R}}}
\def\bSschat{\widehat{\mathbf{\cal S}}}
\def\bTschat{\widehat{\mathbf{\cal T}}}
\def\bUschat{\widehat{\mathbf{\cal U}}}
\def\bVschat{\widehat{\mathbf{\cal V}}}
\def\bWschat{\widehat{\mathbf{\cal W}}}
\def\bXschat{\widehat{\mathbf{\cal X}}}
\def\bYschat{\widehat{\mathbf{\cal Y}}}
\def\bZschat{\widehat{\mathbf{\cal Z}}}
\def\afrak{\mathfrak{a}}
\def\bfrak{\mathfrak{b}}
\def\cfrak{\mathfrak{c}}
\def\dfrak{\mathfrak{d}}
\def\efrak{\mathfrak{e}}
\def\ffrak{\mathfrak{f}}
\def\gfrak{\mathfrak{g}}
\def\hfrak{\mathfrak{h}}
\def\ifrak{\mathfrak{i}}
\def\jfrak{\mathfrak{j}}
\def\kfrak{\mathfrak{k}}
\def\lfrak{\mathfrak{l}}
\def\mfrak{\mathfrak{m}}
\def\nfrak{\mathfrak{n}}
\def\ofrak{\mathfrak{o}}
\def\pfrak{\mathfrak{p}}
\def\qfrak{\mathfrak{q}}
\def\rfrak{\mathfrak{r}}
\def\sfrak{\mathfrak{s}}
\def\tfrak{\mathfrak{t}}
\def\ufrak{\mathfrak{u}}
\def\vfrak{\mathfrak{v}}
\def\wfrak{\mathfrak{w}}
\def\xfrak{\mathfrak{x}}
\def\yfrak{\mathfrak{y}}
\def\zfrak{\mathfrak{z}}
\def\Afrak{\mathfrak{ A}}
\def\Bfrak{\mathfrak{ B}}
\def\Cfrak{\mathfrak{ C}}
\def\Dfrak{\mathfrak{ D}}
\def\Efrak{\mathfrak{ E}}
\def\Ffrak{\mathfrak{ F}}
\def\Gfrak{\mathfrak{ G}}
\def\Hfrak{\mathfrak{ H}}
\def\Ifrak{\mathfrak{ I}}
\def\Jfrak{\mathfrak{ J}}
\def\Kfrak{\mathfrak{ K}}
\def\Lfrak{\mathfrak{ L}}
\def\Mfrak{\mathfrak{ M}}
\def\Nfrak{\mathfrak{ N}}
\def\Ofrak{\mathfrak{ O}}
\def\Pfrak{\mathfrak{ P}}
\def\Qfrak{\mathfrak{ Q}}
\def\Rfrak{\mathfrak{ R}}
\def\Sfrak{\mathfrak{ S}}
\def\Tfrak{\mathfrak{ T}}
\def\Ufrak{\mathfrak{ U}}
\def\Vfrak{\mathfrak{ V}}
\def\Wfrak{\mathfrak{ W}}
\def\Xfrak{\mathfrak{ X}}
\def\Yfrak{\mathfrak{ Y}}
\def\Zfrak{\mathfrak{ Z}}

\def\bAfrak{\mathbf{\mathfrak{A}}}
\def\bBfrak{\mathbf{\mathfrak{B}}}
\def\bCfrak{\mathbf{\mathfrak{C}}}
\def\bDfrak{\mathbf{\mathfrak{D}}}
\def\bEfrak{\mathbf{\mathfrak{E}}}
\def\bFfrak{\mathbf{\mathfrak{F}}}
\def\bGfrak{\mathbf{\mathfrak{G}}}
\def\bHfrak{\mathbf{\mathfrak{H}}}
\def\bIfrak{\mathbf{\mathfrak{I}}}
\def\bJfrak{\mathbf{\mathfrak{J}}}
\def\bKfrak{\mathbf{\mathfrak{K}}}
\def\bLfrak{\mathbf{\mathfrak{L}}}
\def\bMfrak{\mathbf{\mathfrak{M}}}
\def\bNfrak{\mathbf{\mathfrak{N}}}
\def\bOfrak{\mathbf{\mathfrak{O}}}
\def\bPfrak{\mathbf{\mathfrak{P}}}
\def\bQfrak{\mathbf{\mathfrak{Q}}}
\def\bRfrak{\mathbf{\mathfrak{R}}}
\def\bSfrak{\mathbf{\mathfrak{S}}}
\def\bTfrak{\mathbf{\mathfrak{T}}}
\def\bUfrak{\mathbf{\mathfrak{U}}}
\def\bVfrak{\mathbf{\mathfrak{V}}}
\def\bWfrak{\mathbf{\mathfrak{W}}}
\def\bXfrak{\mathbf{\mathfrak{X}}}
\def\bYfrak{\mathbf{\mathfrak{Y}}}
\def\bZfrak{\mathbf{\mathfrak{Z}}}

\def\bAfrakhat{\mathbf{\widehat{\mathfrak{A}}}}
\def\bBfrakhat{\mathbf{\widehat{\mathfrak{B}}}}
\def\bCfrakhat{\mathbf{\widehat{\mathfrak{C}}}}
\def\bDfrakhat{\mathbf{\widehat{\mathfrak{D}}}}
\def\bEfrakhat{\mathbf{\widehat{\mathfrak{E}}}}
\def\bFfrakhat{\mathbf{\widehat{\mathfrak{F}}}}
\def\bGfrakhat{\mathbf{\widehat{\mathfrak{G}}}}
\def\bHfrakhat{\mathbf{\widehat{\mathfrak{H}}}}
\def\bIfrakhat{\mathbf{\widehat{\mathfrak{I}}}}
\def\bJfrakhat{\mathbf{\widehat{\mathfrak{J}}}}
\def\bKfrakhat{\mathbf{\widehat{\mathfrak{K}}}}
\def\bLfrakhat{\mathbf{\widehat{\mathfrak{L}}}}
\def\bMfrakhat{\mathbf{\widehat{\mathfrak{M}}}}
\def\bNfrakhat{\mathbf{\widehat{\mathfrak{N}}}}
\def\bOfrakhat{\mathbf{\widehat{\mathfrak{O}}}}
\def\bPfrakhat{\mathbf{\widehat{\mathfrak{P}}}}
\def\bQfrakhat{\mathbf{\widehat{\mathfrak{Q}}}}
\def\bRfrakhat{\mathbf{\widehat{\mathfrak{R}}}}
\def\bSfrakhat{\mathbf{\widehat{\mathfrak{S}}}}
\def\bTfrakhat{\mathbf{\widehat{\mathfrak{T}}}}
\def\bUfrakhat{\mathbf{\widehat{\mathfrak{U}}}}
\def\bVfrakhat{\mathbf{\widehat{\mathfrak{V}}}}
\def\bWfrakhat{\mathbf{\widehat{\mathfrak{W}}}}
\def\bXfrakhat{\mathbf{\widehat{\mathfrak{X}}}}
\def\bYfrakhat{\mathbf{\widehat{\mathfrak{Y}}}}
\def\bZfrakhat{\mathbf{\widehat{\mathfrak{Z}}}}
%
%
%
%
\def\etal{{\em et al.}}
%
%
%
%
%
\def\cumsum{\mbox{cumsum}}
\def\real{{\mathbb R}}
\def\intinfinf{\int_{-\infty}^{\infty}}
\def\intzinf{\int_{0}^{\infty}}
\def\intzt{\int_0^t}
\def\transpose{{\sf \scriptscriptstyle{T}}}
\def\smhalf{{\textstyle{1\over2}}}
\def\third{{\textstyle{1\over3}}}
\def\twothirds{{\textstyle{2\over3}}}
\def\bell{\bmath{\ell}}
\def\half{\frac{1}{2}}
\def\ninv{n^{-1}}
\def\nhalf{n^{\half}}
\def\mhalf{m^{\half}}
\def\nnhalf{n^{-\half}}
\def\mnhalf{m^{-\half}}
\def\MN{\mbox{MN}}
\def\N{\mbox{N}}
\def\E{\mbox{E}}
\def\pr{P}
\def\var{\mbox{var}}
\def\limn{\lim_{n\to \infty} }
\def\intt{\int_{\tau_a}^{\tau_b}}
\def\sumin{\sum_{i=1}^n}
\def\sumjn{\sum_{j=1}^n}
\def\SUMin{{\displaystyle \sum_{i=1}^n}}
\def\SUMjn{{\displaystyle \sum_{j=1}^n}}
\def\myendthm{\begin{flushright} $\diamond $ \end{flushright}}
\def\convd{\overset{\Dsc}{\longrightarrow}}
\def\convp{\overset{\Psc}{\longrightarrow}}
\def\convas{\overset{a.s.}{\longrightarrow}}
\def\hn{\mbox{H}_0}
\def\ha{\mbox{H}_1}

%
%
%
%
%
\def\trans{^{\transpose}}
\def\inv{^{-1}}
\def\twobyone#1#2{\left[
\begin{array}
{c}
#1\\
#2\\
\end{array}
\right]}
%
%
%
%
%
\def\argmindum{\mathop{\mbox{argmin}}}
\def\argmin#1{\argmindum_{#1}}
\def\argmaxdum{\mathop{\mbox{argmax}}}
\def\argmax#1{\argmaxdum_{#1}}
\def\blockdiag{\mbox{blockdiag}}
\def\corr{\mbox{corr}}
\def\cov{\mbox{cov}}
\def\diag{\mbox{diag}}
\def\dffit{df_{{\rm fit}}}
\def\dfres{df_{{\rm res}}}
\def\dfyhat{df_{\yhat}}
\def\diag{\mbox{diag}}
\def\diagonal{\mbox{diagonal}}
\def\logit{\mbox{logit}}
\def\stdev{\mbox{st.\,dev.}}
\def\stdevhat{{\widehat{\mbox{st.dev}}}}
\def\tr{\mbox{tr}}
\def\trigamma{\mbox{trigamma}}
\def\var{\mbox{var}}
\def\vecof{\mbox{vec}}
\def\AIC{\mbox{AIC}}
\def\AMISE{\mbox{AMISE}}
\def\Corr{\mbox{Corr}}
\def\Cov{\mbox{Cov}}
\def\CV{\mbox{CV}}
\def\GCV{\mbox{GCV}}
\def\LR{\mbox{LR}}
\def\MISE{\mbox{MISE}}
\def\MSSE{\mbox{MSSE}}
\def\ML{\mbox{ML}}
\def\REML{\mbox{REML}}
\def\RMSE{{\rm RMSE}}
\def\RSS{\mbox{RSS}}
\def\Var{\mbox{Var}}
%
%
%
%
\def\bib{\vskip12pt\par\noindent\hangindent=1 true cm\hangafter=1}
\def\jump{\vskip3mm\noindent}
\def\mybox#1{\vskip1mm \begin{center}
        \hspace{.0\textwidth}\vbox{\hrule\hbox{\vrule\kern6pt
\parbox{.9\textwidth}{\kern6pt#1\vskip6pt}\kern6pt\vrule}\hrule}
        \end{center} \vskip-5mm}
\def\lboxit#1{\vbox{\hrule\hbox{\vrule\kern6pt
      \vbox{\kern6pt#1\vskip6pt}\kern6pt\vrule}\hrule}}
\def\boxit#1{\begin{center}\fbox{#1}\end{center}}
\def\thickboxit#1{\vbox{{\hrule height 1mm}\hbox{{\vrule width 1mm}\kern6pt
          \vbox{\kern6pt#1\kern6pt}\kern6pt{\vrule width 1mm}}
               {\hrule height 1mm}}}
\def\instep{\vskip12pt\par\hangindent=30 true mm\hangafter=1}
\def\uWand{\underline{Wand}}
\def\remtask#1#2{\mmnote{\thickboxit
                 {\bf #1\ \theremtask}}\refstepcounter{remtask}}
%
%
%

%
%
\def\aism{{\it Ann. Inst. Statist. Math.}\ }
\def\ajs{{\it Austral. J. Statist.}\ }
\def\ANNSTAT{{\it The Annals of Statistics}\ }
\def\annmath{{\it Ann. Math. Statist.}\ }
\def\applstat{{\it Appl. Statist.}\ }
\def\BIOMETRICS{{\it Biometrics}\ }
\def\cjs{{\it Canad. J. Statist.}\ }
\def\csda{{\it Comp. Statist. Data Anal.}\ }
\def\cstm{{\it Comm. Statist. Theory Meth.}\ }
\def\ieeetit{{\it IEEE Trans. Inf. Theory}\ }
\def\isr{{\it Internat. Statist. Rev.}\ }
\def\JASA{{\it Journal of the American Statistical Association}\ }
\def\JCGS{{\it Journal of Computational and Graphical Statistics}\ }
\def\jscs{{\it J. Statist. Comput. Simulation}\ }
\def\jma{{\it J. Multivariate Anal.}\ }
\def\jns{{\it J. Nonparametric Statist.}\ }
\def\JRSSA{{\it Journal of the Royal Statistics Society, Series A}\ }
\def\JRSSB{{\it Journal of the Royal Statistics Society, Series B}\ }
\def\JRSSC{{\it Journal of the Royal Statistics Society, Series C}\ }
\def\jspi{{\it J. Statist. Planning Inference}\ }
\def\ptrf{{\it Probab. Theory Rel. Fields}\ }
\def\sankhyaa{{\it Sankhy$\bar{{\it a}}$} Ser. A\ }
\def\sjs{{\it Scand. J. Statist.}\ }
\def\spl{{\it Statist. Probab. Lett.}\ }
\def\statsci{{\it Statist. Sci.}\ }
\def\techno{{\it Technometrics}\ }
\def\tpa{{\it Theory Probab. Appl.}\ }
\def\zw{{\it Z. Wahr. ver. Geb.}\ }
%
%
%
%
\def\Brent{{\bf BRENT:}\ }
\def\David{{\bf DAVID:}\ }
\def\Erin{{\bf ERIN:}}
\def\Gerda{{\bf GERDA:}\ }
\def\Joel{{\bf JOEL:}\ }
\def\Marc{{\bf MARC:}\ }
\def\Matt{{\bf MATT:}\ }
\def\Tianxi{{\bf TIANXI:}\ }
%
%
%
%
\def\bZE{\bZ_{\scriptscriptstyle E}}
\def\bZT{\bZ_{\scriptscriptstyle T}}
\def\bbE{\bb_{\scriptscriptstyle E}}
\def\bbT{\bb_{\scriptscriptstyle T}}
\def\bbhatT{\bbhat_{\scriptscriptstyle T}}
\def\fX{f_{\scriptscriptstyle X}}
\def\sigeps{\sigma_{\varepsilon}}
\def\bVtheta{\bV_{\smbtheta}}
\def\bVthetainv{\bVtheta^{-1}}
\def\bKsc{\boldsymbol{\Ksc}}
\def\bxbar{\bar{\bx}}
\def\bPL{b^{\scriptscriptstyle{\rm PL}}}
\def\bVA{b^{\scriptscriptstyle{\rm VA}}}
\def\zPL{z^{\scriptscriptstyle{\rm PL}}}
\def\zVA{z^{\scriptscriptstyle{\rm VA}}}
\def\bYmis{\bY_{\scriptscriptstyle{\rm mis}}}
\def\bYmishat{{\widehat{\bYmis}}}
\def\bYmisone{\bY_{\scriptscriptstyle{\rm mis,1}}}
\def\bYmistwo{\bY_{\scriptscriptstyle{\rm mis,2}}}
\def\bYobs{\bY_{\scriptscriptstyle{\rm obs}}}
\def\bdobs{\bd_{\scriptscriptstyle{\rm obs}}}
\def\bdmis{\bd_{\scriptscriptstyle{\rm mis}}}
%
%
%
%
\def\bfDelta{{\mbox{\boldmath$\Delta$}}}
\def\bfkappa{{\mbox{\boldmath$\kappa$}}}
\def\bfgamma{{\mbox{\boldmath$\gamma$}}}
\def\bftheta{{\mbox{\boldmath$\theta$}}}
\def\bfmu{{\mbox{\boldmath$\mu$}}}
\def\bfdelta{{\mbox{\boldmath$\delta$}}}
\def\bfeps{{\mbox{\boldmath$\varepsilon$}}}
\def\bfnu{{\mbox{\boldmath$\nu$}}}
\def\bfzeta{{\mbox{\boldmath$\zeta$}}}
\def\bfchi{{\mbox{\boldmath$\chi$}}}
\def\bbX{\mathbb{X}}
\def\bbV{\mathbb{V}} 
\def\bbA{\mathbb{A}}
\def\bbB{\mathbb{B}}
\def\bbK{\mathbb{K}}
\def\bbP{\mathbb{P}}
\def\bbD{\mathbb{D}}

\def\Abb{\mathbb{A}}
\def\Bbb{\mathbb{B}}
\def\Cbb{\mathbb{C}}
\def\Dbb{\mathbb{D}}
\def\Ebb{\mathbb{E}}
\def\Fbb{\mathbb{F}}
\def\Gbb{\mathbb{G}}
\def\Hbb{\mathbb{H}}
\def\Ibb{\mathbb{I}}
\def\Jbb{\mathbb{J}}
\def\Kbb{\mathbb{K}}
\def\Lbb{\mathbb{L}}
\def\Mbb{\mathbb{M}}
\def\Nbb{\mathbb{N}}
\def\Mbb{\mathbb{M}}
\def\Nbb{\mathbb{N}}
\def\Obb{\mathbb{O}}
\def\Pbb{\mathbb{P}}
\def\Qbb{\mathbb{Q}}
\def\Rbb{\mathbb{R}}
\def\Sbb{\mathbb{S}}
\def\Tbb{\mathbb{T}}
\def\Ubb{\mathbb{U}}
\def\Vbb{\mathbb{V}}
\def\Wbb{\mathbb{W}}
\def\Xbb{\mathbb{X}}
\def\Ybb{\mathbb{Y}}
\def\Zbb{\mathbb{Z}}

\def\Abbtilde{\widetilde{\mathbb{A}}}
\def\Bbbtilde{\widetilde{\mathbb{B}}}
\def\Cbbtilde{\widetilde{\mathbb{C}}}
\def\Dbbtilde{\widetilde{\mathbb{D}}}
\def\Ebbtilde{\widetilde{\mathbb{E}}}
\def\Fbbtilde{\widetilde{\mathbb{F}}}
\def\Gbbtilde{\widetilde{\mathbb{G}}}
\def\Hbbtilde{\widetilde{\mathbb{H}}}
\def\Ibbtilde{\widetilde{\mathbb{I}}}
\def\Jbbtilde{\widetilde{\mathbb{J}}}
\def\Kbbtilde{\widetilde{\mathbb{K}}}
\def\Lbbtilde{\widetilde{\mathbb{L}}}
\def\Mbbtilde{\widetilde{\mathbb{M}}}
\def\Nbbtilde{\widetilde{\mathbb{N}}}
\def\Mbbtilde{\widetilde{\mathbb{M}}}
\def\Nbbtilde{\widetilde{\mathbb{N}}}
\def\Obbtilde{\widetilde{\mathbb{O}}}
\def\Pbbtilde{\widetilde{\mathbb{P}}}
\def\Qbbtilde{\widetilde{\mathbb{Q}}}
\def\Rbbtilde{\widetilde{\mathbb{R}}}
\def\Sbbtilde{\widetilde{\mathbb{S}}}
\def\Tbbtilde{\widetilde{\mathbb{T}}}
\def\Ubbtilde{\widetilde{\mathbb{U}}}
\def\Vbbtilde{\widetilde{\mathbb{V}}}
\def\Wbbtilde{\widetilde{\mathbb{W}}}
\def\Xbbtilde{\widetilde{\mathbb{X}}}
\def\Ybbtilde{\widetilde{\mathbb{Y}}}
\def\Zbbtilde{\widetilde{\mathbb{Z}}}

%
%
%
%
\def\miss{\mbox{{\tiny miss}}}
\def\obs{\scriptsize{\mbox{obs}}}

%
%
%
%
\def\bmath#1{\mbox{\boldmath$#1$}}
\def\fat#1{\hbox{\rlap{$#1$}\kern0.25pt\rlap{$#1$}\kern0.25pt$#1$}}
\def\wh{\widehat}
\def\flambda{\fat{\lambda}}
\def\beps{\bmath{\varepsilon}}
\def\bSlambda{\bS_{\lambda}}
\def\ErrorSS{\mbox{RSS}}
\def\bsqbar{\bar{{b^2}}}
\def\bcubar{\bar{{b^3}}}
\def\plargest{p_{\rm \,largest}}
\def\summheading#1{\subsection*{#1}\hskip3mm}
\def\summbreak{\vskip3mm\par}
\def\df{df}
\def\adf{adf}
\def\dffit{df_{{\rm fit}}}
\def\dfres{df_{{\rm res}}}
\def\dfyhat{df_{\yhat}}
\def\sigb{\sigma_b}
\def\sigu{\sigma_u}
\def\sigepshat{{\widehat\sigma}_{\varepsilon}}
\def\siguhat{{\widehat\sigma}_u}
\def\sigepshat{{\widehat\sigma}_{\varepsilon}}
\def\sigbhat{{\widehat\sigma}_b}
\def\sighat{{\widehat\sigma}}
\def\sigsqb{\sigma^2_b}
\def\sigsqeps{\sigma^2_{\varepsilon}}
\def\sigsqepszerohat{{\widehat\sigma}^2_{\varepsilon,0}}
\def\sigsqepshat{{\widehat\sigma}^2_{\varepsilon}}
\def\sigsqbhat{{\widehat\sigma}^2_b}
\def\dfnumer{{\rm df(II}|{\rm I)}}
\def\mhatlam{{\widehat m}_{\lambda}}
\def\calD{\Dsc}
\def\Aeps{A_{\epsilon}}
\def\Beps{B_{\epsilon}}
\def\Ab{A_b}
\def\Bb{B_b}
\def\bXtmain{\tilde{\bX}_r}
\def\main{\mbox{\tt main}}
\def\argminbetab{\argmin{\bbeta,\bb}}
\def\calB{\Bsc}
\def\respvar{\mbox{\tt log(amt)}}

\def\Abb{\mathbb{A}}
\def\Zbb{\mathbb{Z}}
\def\Wbb{\mathbb{W}}
\def\Wbbhat{\widehat{\mathbb{W}}}
\def\Kbbtilde{\widetilde{\mathbb{K}}}
\def\Pbbtilde{\widetilde{\mathbb{P}}}
\def\Dbbtilde{\widetilde{\mathbb{D}}}
\def\Bbbtilde{\widetilde{\mathbb{B}}}

\def\Abbhat{\widehat{\mathbb{A}}}

\def\ellhat{\widehat{\ell}}
\def\pn{\phantom{-}}
\def\pp{\phantom{1}}

\def\PP{\stackrel{P}{\rightarrow}}
\def\DD{\Rightarrow}
%
%
\newcommand{\bXvec}{\vec{\bX}}
\newcommand{\bZvec}{\vec{\bZ}}
\newcommand{\bXveci}{\vec{\bX}_{i}}
\def\pibar{\bar{\pi}}

\maketitle
\thispagestyle{empty}


\setstretch{1}
\begin{abstract}
Meta-analysis aggregates information across related studies to provide more reliable statistical inference and has been a vital tool for assessing the safety and efficacy of many high profile pharmaceutical products.  A key challenge in conducting a meta-analysis is that the number of related studies is typically small. Applying classical methods that are asymptotic in the number of studies can compromise the validity of inference, particularly when heterogeneity across studies is present.  Moreover, serious adverse events are often rare and can result in one or more studies with no events in at least one study arm.  Practitioners often apply arbitrary continuity corrections or remove zero-event studies to stabilize or define effect estimates in such settings, which can further invalidate subsequent inference.  To address these significant practical issues, we introduce an exact inference method for comparing event rates in two treatment arms under a random effects framework, which we coin ``XRRmeta''.  In contrast to existing methods, the coverage of the confidence interval from XRRmeta is guaranteed to be at or above the nominal level (up to Monte Carlo error) when the event rates, number of studies, and/or the within-study sample sizes are small. Extensive numerical studies indicate that XRRmeta does not yield overly conservative inference and we apply our proposed method to two real-data examples using our open source \texttt{R} package.\\
\end{abstract}

\noindent{\bf{Key Words:}} Exact Inference; Meta-analysis; Random effects model; Rare events; Rosiglitazone

\maketitle

\setstretch{1.5}
\section{Introduction}
\label{s:intro}
Meta-analysis is widely used in clinical research to aggregate information from similar studies to yield more efficient inference and improve statistical power \supercite{egger1997meta, egger1997meta2, normand1999tutorial}.  It is particularly useful for assessing the frequency of adverse events in drug safety studies as single studies are typically powered to establish treatment efficacy and adverse events are rare. However, the validity of most existing meta-analytic approaches rests on the asymptotic distribution of the combined point estimator, which can be unreliable when any of the following conditions hold: (i) the event rates are low, (ii) the number of studies is not large, and (iii) the study-specific sample sizes are small \supercite{brown2002confidence, j2004add, bradburn2007much, cai2010meta}.  This is a significant practical issue as these conditions are common within the literature.  A study of 500 Cochrane systematic reviews found that 50\% of drug safety meta-analyses contained an outcome with a rare event rate ($<5\%$) and 30\% contained at least one study with no events in one arm \supercite{vandermeer2009meta}.  Additionally, only 16\% of meta-analyses considered had 4 or more studies.  A well-known and controversial example where these issues arise involves the type II diabetes drug, rosiglitazone, which was suspected to increase the risk of myocardial infarction (MI) and cardiovascular death (CVD)\supercite{nissen2007effect}.  The available data are presented in Supplementary Table S1 and provided an initial motivation for this work.

In the original meta-analysis of the rosiglitazone data, the authors utilized the conventional fixed effect Peto method based on the combined odds ratio \supercite{nissen2010rosiglitazone}.  Because the events of interest were extremely rare, 25 of the 48 studies had no CVD and 10 studies had no MIs.  The authors excluded these double-zero (DZ) studies from their analysis.  Much discussion has since ensued regarding conflicting conclusions of alternative analyses of the rosiglitazone data based on the widely used fixed-effect Mantel-Haenszel method, which similarly relies on a normality approximation and requires removal of DZ studies or use of continuity corrections \supercite{liebson2007rosiglitazone, shuster2007fixed, friedrich2009rosiglitazone, cai2010meta, nissen2010rosiglitazone, claggett2011analytical, efthimiou2018practical}.  As there is no clear guidance for removing studies or applying continuity corrections, \cite{tian2008exact} proposed an exact confidence interval for fixed-effect meta-analyses, based on the combination of study-specific exact confidence intervals, which utilizes all available data and avoids continuity corrections \supercite{sankey1996assessment, bradburn2007much}. \cite{liu2014exact} later extended this approach with a method that combines $p$-value functions using the mid-$p$ adaption of Fisher’s exact method.  While these procedures are more robust than their classical counterparts, the underlying fixed effect assumption implies that all study-specific treatment effects are identical.  In the case of the rosiglitazone study, for example, this assumption is likely violated as the studies had different eligibility criteria, medication doses, control and concomitant medications, and  follow-up times \supercite{shuster2007fixed}.

An alternative and less restrictive approach is to employ a random effects analysis \supercite{beisemann2020comparison, gunhan2020random, jansen2022random}. The most popular procedure is the DerSimonian-Laird (DL) method \supercite{dersimonian1986meta}.  The DL combined point estimator is a linear combination of study-specific estimates of the effect of interest with weights based on the within- and between-study variation estimates.  Depending on the chosen effect measure, similar issues arise with respect to continuity corrections or removal of DZ studies, making the application of the DL method to the rosiglitazone data questionable \supercite{friedrich2007inclusion, bhaumik2012meta, cheng2016impact}.  Moreover,  the DL method relies on the assumption that the study-specific effects follow a normal distribution,  which is unlikely to hold in the rare events setting.  The validity of inference based on the DL method is further threatened when the number of studies is small as the between-study variance is imprecisely estimated \supercite{langan2019comparison}.  Difficulties in estimating the between-study variance has similarly limited the application of random effects regression-based approaches, particularly in the rare event setting \supercite{veroniki2016methods}.  To overcome these challenges, Shuster et al.\ introduced a ratio estimator for random effects meta-analysis for the setting of low event rates leveraging results from sampling theory \supercite{shuster2007fixed}.  Their findings differed from the original meta-analysis of the rosiglitazone data, indicating an elevated risk of CVD and no increased risk of MI with the use of the medication.  Cai  et al.\ later developed a likelihood-based approach based on a Poisson random effects model \supercite{cai2010meta}.  In contrast to the previously proposed ratio estimator, the authors employed a conditional inference argument to avoid continuity corrections and theoretically justified exclusion of DZ studies.  Jiang et al.\ also introduced a method for obtaining profile confidence limits for the risk difference using importance sampling \supercite{jiang2020accurate}.  While these methods all target the setting with low event rates, the proposed inference procedures remain asymptotic in the number of studies and, to the best of our knowledge, are not available in open-source software.  More recently,  Zabriskie et al.\ proposed a permutation-based approach based on conditional logistic regression \supercite{zabriskie2021permutation}.  However, this method cannot be applied to all data sets, does not uniformly guarantee type I error control, and is computationally intensive.  

To address the limitations of existing methods, we introduce an e\underline{x}act inference procedure for \underline{r}andom effects meta-analysis of a treatment effect in the two-sample setting with \underline{r}are events, which we coin ``XRRmeta''. XRRmeta is particularly attractive for rare event outcome data as it is based on a conditional inference argument that justifies the removal of DZ studies.  Moreover, XRRmeta yields a confidence interval (CI) through inversion of exact tests and is therefore guaranteed to achieve coverage at or above the nominal level (up to Monte Carlo error).  Importantly, our numerical studies indicate that our choice of test statistic yields inference that is not overly conservative and enables us to develop a procedure that is computationally feasible to run a personal laptop computer.  XRRmeta is also available in open-source \texttt{R} software at \href{https://github.com/zrmacc/RareEventsMeta}{https://github.com/zrmacc/RareEventsMeta} to encourage use in practice.

The remainder of this paper is organized as follows.  In Sections \ref{sec: prob} and \label{sec: methods}, we present the methodological and computational details of XRRmeta.  The performance of XRRmeta is then evaluated with extensive simulation studies in Section \ref{sec: sim}.  In Section \ref{sec: real}, we apply our procedure to the rosiglitazone study and a recent meta-analysis of face mask use in preventing person-to-person transmission of severe acute respiratory syndrome coronavirus 2 (SARS-CoV-2) and coronavirus disease 2019 (COVID-19).  We close with additional remarks and avenues for future research in Section \ref{sec: discussion}.


\section{Problem Setup}\label{sec: prob}
\subsection{Notations and Assumptions}\label{sec: data}
Our goal is to compare the rates of an event of interest from multiple studies comparing the same treatments.  The observed data consists of 
$$\Dsc^0 = \{ (Y_{ij}, N_{ij}) \mid i = 1, \dots K_{tot} \text{ and } j = 1, 2 \}$$
 where $Y_{ij}$ is the number of events out of $N_{ij}$ subjects in the $i$th study  and $j$th treatment arm and $K_{tot}$ is the total number of studies. Without loss of generality, we let arm 1 correspond to the treated group and arm 2 correspond to the control group.  As the $Y_{ij}$ are counts, we assume that they follow a Poisson distribution with the rate parameter following a log-linear model.  That is,
$$Y_{ij} \overset{\text{ind}}{\sim} \text{Poisson}( N_{ij} \lambda_{ij})  \quad \mbox{where} \quad  \lambda_{ij} = \lambda_{i2} e^{X_{ij} \xi_i},$$
$X_{ij} =\mathbb{I}(j = 1)$ is a binary treatment indicator, $\lambda_{ij}$ is the event rate in the $j$th treatment arm of the $i$th study, and $\xi_i = \log( \lambda_{i1}/\lambda_{i2})$ is the log relative risk. 

Under the null hypothesis of no treatment effect (i.e., $\xi_i =0$), a sufficient statistic for the nuisance parameter $\lambda_{i2}$ is $Y_{i \bcdot} = Y_{i1} + Y_{i2}$, the total number of events in the $i$th study.  We take the classical approach of basing inference on the conditional distribution of $Y_{i1}$ given $Y_{i \bcdot}$ to eliminate the nuisance parameter $\lambda_{i2}$ \supercite{cai2010meta}.  Simple calculations show that
\begin{equation}
  Y_{i1} \mid Y_{i \bcdot}  \sim \text{Binomial}\left\{ Y_{i \bcdot} , \text{expit}(\xi_i + S_i) \right\}  \label{cond_dist}
\end{equation}
where $S_i = \log(N_{i1}/N_{i2})$ and $\text{expit}(x) = 1/(1+e^{-x})$.  As we are primarily interested in the rare event rate setting, it is important to note that basing inference on \ref{cond_dist} justifies the exclusion of DZ studies from analysis as they do not provide information on the relative risk, $\exp(\xi_i)$.  This enables use to utilize the $K \le K_{tot}$ non-DZ studies for analysis \supercite{bohning2021identity}.   

\subsection{Parameter of Interest}
To assess the relative event rates in the two treatment arms, we define a treatment contrast
$$\pi_i \equiv \text{expit}(\xi_i) =  \frac{\lambda_{i1}}{\lambda_{i1} + \lambda_{i2}}$$
which measures the magnitude of the event rate in the treated group relative to the cumulative event rate across both treatment arms in the $i$th study. The setting of no treatment effect corresponds to $\pi_{i} = 0.5$ while $\pi_{i} > 0.5$ indicates the event is more common in the $i$th study's treated arm.  We further assume that $\pi_{i}$ is a random effect with
$$\pi_i   \sim \text{Beta}(\alpha_0, \beta_0) \quad \mbox{with} \quad \alpha_0, \beta_0 > 1$$   
 to account for between-study heterogeneity.  Under the Beta random effects distribution for $\pi_i$, it follows that
 \begin{align*}
\E( \pi_i ) = \frac{\alpha_0}{\alpha_0 + \beta_0} = \mu_0 \quad \mbox{and} \quad  \Var(\pi_i  ) = \mu_0 (1-\mu_0) \tau_0  = \nu_0
\end{align*}
 where $\tau_0 = (\alpha_0 + \beta_0 + 1)^{-1}$ quantifies the between-study variability.  Under the balanced design with equal sample sizes in both arms, the corresponding random effects model for $Y_{i1} \mid Y_{i \bcdot}$ simplifies to the familiar Beta Binomial (BB) model.  In the subsequent sections, we develop an exact method to make inference on $\mu_0$.  We reparameterize the distribution of $\pi_i$ with respect to $\mu_0$ and $\nu_0$ for clarity of presentation with the following equalities
 \begin{equation*} 
 \alpha_0 = \mu_0 \left\{\frac{\mu_0(1-\mu_0) - \nu_0}{\nu_0}  \right\} \quad \mbox{and} \quad \beta_0 = (1-\mu_0) \left\{ \frac{\mu_0(1-\mu_0) - \nu_0}{\nu_0}  \right\}.
 \end{equation*}

Before introducing our proposal, we note that an alternative approach is to base inference on $E(\xi_i)$ with $\xi_i$ following a normal distribution.  However, it is challenging to appropriately specify the standard deviation in the setting of rare events \supercite{fleiss1993review}.  To clarify this point, consider the following toy example.  Suppose we observe data from 6 studies with a balanced design with study-specific sample sizes of 100 and  
$$\{ (Y_{i1}, Y_{i2}) = (0, 20) \mid  i = 1,\dots, 6 \}$$
Intuition suggests that the treatment is protective. However, by placing a normal random effects distribution on the log relative risk, one cannot rule out the possibility that $\xi_i \sim N(10, 1000).$  Under this model, there is approximately $2^{-6}=1.5\%$ probability of obtaining the observed data by chance, which is surprisingly large given the intuition that $E(\xi_i)$ should be very negative.  It is unclear how to naturally constrain the standard deviation of the normal distribution to prevent this phenomena from occurring.  

In contrast to basing inference on $E(\xi_i)$, our choice of random effects distribution affords several benefits \supercite{gronsbell2018}.  Our requirement on the support of $(\alpha_0, \beta_0)$ implies that
\begin{align*}
 \nu_0  \le \mu_0(1-\mu_0) \min \left( \frac{\mu_0}{1+\mu_0}, \frac{1-\mu_0}{2-\mu_0} \right)=\nu_{\sup}(\mu_0).
 \end{align*}
This constraint reduces the parameter space of the standard beta distribution defined with $(\alpha_0, \beta_0) > 0$ to $\left\{(\mu_0, \nu_0) \mid \nu_0\le \nu_{\sup}(\mu_0) \right\}$.  This assumption guarantees that the random effects distribution is unimodal so that we may (i) appropriately interpret $\mu_0$ as the center of $\pi_i$ and (ii) ensure that $\mu_0$ is identifiable.  Going back to our toy example, one would expect that $\mu_0$ is close to zero.  However, with $\pi_i\sim \text{Beta}(0.001, 0.004)$ the corresponding $\mu_0 = 1/5$.  Under this random effects distribution, there is approximately 26\% probability that all the $\pi_i \approx 0 $ by chance.  This paradox arises from the fact that the $\text{Beta}(0.001, 0.004)$ distribution places 4/5 and 1/5 probability on $\pi_i=0$ and $\pi_i=1$, respectively.  Our constraint naturally eliminates bimodal prior distributions to rectify this behavior. 

{ \paragraph{Remark 1} 
While we focus on meta-analysis of 2$\times$2 tables, our framework also applies to the analysis of incidence rates.  In this setting, the observable data consists of $\{ (Y_{ij}, T_{ij})  \mid i = 1, \dots K_{tot}, j = 1,2 \}$ where $Y_{ij}$ is the number of events in the total follow-up time $T_{ij}$  in the $i$th study and $j$th treatment arm.  Inference may proceed with analogous assumptions on $Y_{i1} \mid  Y_{i \bcdot}$ as those stated in Section \ref{sec: prob}.}

\section{Methods}\label{sec: method}
\subsection{Exact Inference Procedure}
While basing inference on the distribution of $Y_{i1} \mid Y_{i \bcdot}$ eliminates the nuisance parameter dictating the event rate in the control arm, making inference on $\mu_0$ demands consideration of $\nu_0$.  To develop a confidence interval (CI), we propose to invert unconditional tests with respect to $\nu_0$.  More specifically, we perform a test of the null hypothesis
$$H_0 : \mu_0 = \mu$$  
based on a test statistic $T(\mu; \Dsc^0)$ that is function of both $\mu$ and the observed data $\Dsc^0$. We detail the choice of $T(\mu; \Dsc^0)$ in Section \ref{sec: teststat} and here only assume that larger values of the test statistic lead to a rejection of $H_0$.  The unconditional test eliminates $\nu_0$ with the profile $p$ value defined as
\begin{equation}\label{eq:pval}
p(\mu; \Dsc^0) =  \sup_{\nu} P \left\{ T(\mu; \Dsc^{\mu, \nu} ) \ge  T(\mu; \Dsc^0) \mid \Dsc^0 \right\} = \sup_{\nu} p(\mu, \nu; \Dsc^0).
\end{equation}
The probability in \ref{eq:pval} is taken with respect to data, $\Dsc^{\mu, \nu}$, following the random effects model with parameters $\mu$ and $\nu$ outlined in Section \ref{sec: data}.  For an $\alpha$-level test, the null hypothesis is rejected when $p(\mu; \Dsc^0) < \alpha$ and the corresponding $(1-\alpha) 100\%$ CI includes all $\mu$ such that $p(\mu; \Dsc^0) \ge \alpha$.  Our proposed procedure, XRRmeta, utilizes this framework to yield an exact CI as detailed in Figure \ref{figure:method}.    

\begin{figure}[H]
\begin{tcolorbox}[title={Overview of XRRmeta}]
\begin{description}
       \item[Step 1.] Let $G = [\mu_{L}, \mu_{U}] \times [\nu_{L}, \nu_{U}]$ by a dense $H \times J$ grid covering the true values of $\mu$ and $\nu$.  For each $(\mu, \nu) \in G$ compute the $p$ value  \\
$$p(\mu, \nu; \Dsc^0) = P \left\{ T(\mu; \Dsc^{\mu, \nu} ) \ge  T(\mu; \Dsc^0) \mid \Dsc^0 \right\}.$$ \\
\item[Step 2.] 
Project the $(1-\alpha)100\%$ confidence region for $(\mu_0, \nu_0)$ given by\\
$$\Omega_{1-\alpha}(\Dsc^0) = \{ (\mu, \nu) \mid p(\mu, \nu; \Dsc^0)  \ge \alpha \}$$ \\
to the $\mu$ axis to obtain the $(1-\alpha)100\%$ CI for $\mu_0$ as\\
$$\left\{ \inf_{\mu \in [\mu_{L}, \mu_{U}]} \Omega_{1-\alpha}(\Dsc^0),  \sup_{\mu \in [\mu_{L}, \mu_{U}]}\Omega_{1-\alpha}(\Dsc^0)  \right\}.$$ 
\end{description}
\end{tcolorbox}
\caption{{{Two steps required to obtain an exact confidence interval (CI) with XRRmeta.}}}
\label{figure:method}
\end{figure}

The primary complication in implementing XRRmeta is that the cumulative distribution function of $T(\mu; \Dsc^{\mu, \nu})$ is required to calculate $p(\mu, \nu; \Dsc^0)$.  As the distribution function is likely unavailable in analytic form, an approximation can be obtained with Monte Carlo (MC) simulation.  More specifically, for a large number $M$, we can repeat the following steps for $m = 1, \dots, M$ within Step 1 of the XRRmeta method in Figure \ref{figure:method}: 
\begin{description}
\item[Step 1a.]  Generate $\Dsc_m = \{ (Y_{i1}^m,  Y_{i2}^m) \mid i = 1, \dots, K \}$ where $Y_{i2}^m = Y_{i \bcdot} - Y_{i1}^m$,
$$  Y_{i1}^m  \sim  \text{Binom}\big[ Y_{i \bcdot} , \text{expit}\{\text{logit}(\pi_{i}) + S_i\}\big], \quad \pi_{i} \sim \text{Beta}(\mu, \nu), \quad \mbox{\&} \quad S_i = \log(N_{i1}/N_{i2}).$$
\item[Step 1b.] Compute $T_m = T(\mu; \Dsc_{m})$.
\end{description}
We may then calculate $p(\mu, \nu; \Dsc^0)$ with $M^{-1} \sum_{m = 1}^M I\{T_m \ge T(\mu; \Dsc^0)\}$.  However, the MC procedure must be executed $HJ$ times, which is computationally burdensome with a large number of replications. We therefore strategically design $T(\mu; \Dsc^0)$ to allow for fast computation of the exact CI.  To motivate our proposal, we begin by introducing the test statistic under a balanced design in which the underlying random effects model reduces to the BB model with parameters $\mu_0$ and $\nu_0$.  We then describe an augmentation of the proposed test statistic under the more likely scenario of an unbalanced design and further simplifications to accelerate computation.

\subsection{Test Statistic}\label{sec: teststat}
\subsubsection{Balanced Design}\label{sec: balan}
Here we consider the balanced design setting with $N_{i1} = N_{i2}$ for $i = 1, \dots, K$ so that $Y_{i1}\mid Y_{i \bcdot}  \sim \text{BB}\left( Y_{i \bcdot}, \mu_0, \nu_0 \right)$.  Natural choices of $T(\mu; \Dsc^0)$ include the Wald, score, and likelihood ratio statistic. We propose a Wald statistic using method of moments estimators for $\mu_0$ and $\nu_0$ as they do not require iterative calculations that would substantially increase the computational time of the MC procedure of XRRmeta. To this end, note that under the balanced design 
\begin{align*}
\E\left(\frac{Y_{i1}}{Y_{i\cdot}}\right) =  \mu_0 \quad \mbox{and} \quad  \E \left\{ \left(\frac{Y_{i1}}{Y_{i\cdot}}\right)^2 \right\} = \left(1- \frac{1}{Y_{i\cdot}}\right) (\mu_0^2 + \nu_0) +  \frac{\mu_0}{Y_{i\cdot}}.
\end{align*}
We respectively obtain method of moment estimators for $\mu_0$ and $\nu_0$ as
\begin{equation}\label{eq: firstMOM}
\muhat = K^{-1} \sum_{i = 1}^K \frac{Y_{i1}}{Y_{i\cdot}}  \quad \mbox{and} \quad
\nuhat =  \max \left\{ 0,   \frac{ \sum_{i = 1}^K  \left\{ \left( \frac{\Ytilde_{i1}} {\Ytilde_{i\cdot}} \right)^2 -   \frac{\muhat_{int} }{\Ytilde_{i\cdot}} \right\}  }{\sum_{i = 1}^K  ( 1 - \frac{1} {\Ytilde_{i\cdot}} ) }- \muhat_{int}^2 \right\} 
\end{equation}
where $\muhat_{int} = K^{-1}\sum_{i = 1}^K \frac{\Ytilde_{i1}} {\Ytilde_{i\cdot}}$ and $(\Ytilde_{i1}, \Ytilde_{i\cdot}) = (Y_{i1} +  0.5 , Y_{i\cdot} +  1).$
The continuity correction is utilized for estimating $\nu_0$ so that (i) $\Var \left(\frac{\Ytilde_{i1}} {\Ytilde_{i\cdot}} \mid \pi_i \right)  > 0$ when $Y_{i1} = 0$ or $Y_{i1} = Y_{i\cdot}$ in all $K$ studies and (ii) all studies contribute to estimation as it is possible that $Y_{i\cdot} = 1$ in the rare event setting \supercite{bradburn2007much}. In contrast to existing procedures, this correction does not impact the validity of XRRmeta as it is based on the exact distribution of the test statistic. The Wald test statistic is simply 
\begin{equation*}\label{eq:TS}
T(\mu; \Dsc^0) = \frac{(\muhat - \mu)^2}{ \widehat{\text{Var}}(\muhat )} \quad \mbox{where} \quad \widehat{\text{Var}}(\muhat ) = K^{-2} \left\{ \sum_{i = 1}^K \frac{\muhat(1-\muhat)}{\Ytilde_{i\cdot}} + \left(1- \frac{1}{\Ytilde_{i\cdot}}\right)\nuhat   \right\}.
\end{equation*}

It is important to note that the choice of test statistic for XRRmeta is not unique.  In additional to computational efficiency, it is necessary to consider the impact of the test statistic on statistical efficiency and hence the length of the resulting CI.  For example, one may also estimate $\mu_0$ with $\sum_{i=1}^K Y_{i1}/\sum_{i=1}^K Y_{i\cdot},$ which is expected to be more accurate in the presence of low between-study heterogeneity.  While the inverse variance estimator is generally expected to provide the best performance, this is true only when the sample size (i.e., the total number of events) of every individual study is sufficiently large.  We initially investigated the inverse variance estimator of $\mu_{0}$ in our numerical studies, but did not observe a substantial improvement in the efficiency of XRRmeta relative to our proposed method of moments estimators. 

\subsubsection{Unbalanced Design}\label{sec: unbalan}
Building on Section \ref{sec: balan}, we next consider the more realistic setting of an unbalanced design in which the treated and control arms have different sample sizes for at least one study.  In this setting, $Y_{i1} \mid Y_{i \bcdot}$ no longer follows the familiar BB model.  The first two moments are $\E\left(\frac{Y_{i1}}{Y_{i\cdot}}\right) = \E \left\{ \text{expit}(\xi_i + S_i)  \right\}$ and
\begin{align*}
  \E \left\{ \left(\frac{Y_{i1}}{Y_{i\cdot}}\right)^2 \right\} = \E \left[ \frac{\text{expit}(\xi_i + S_i)}{ Y_{i\cdot}}  + \left(1- \frac{1}{Y_{i\cdot}}\right) \left\{\text{expit}(\xi_i + S_i)\right\}^2 \right ].
\end{align*}
The previously proposed method of moment estimators from the balanced design setting therefore cannot be directly employed.  To avoid estimators requiring iterative calculation, we propose weighted counterparts of our previous proposals motivated by ``resampling'' a subset of the larger arm to mimic the balanced design setting.  That is, for each study the possible outcomes for the event of interest are enumerated under a balanced design and assigned a weight according to their likelihood conditional on the observed data.  The moment estimators are then calculated with the enumerated data and corresponding weights.

For exposition, we detail the case with $N_{i1} > N_{i2}$ for $i = 1, \dots, K$.  Analogous results hold for the setting with smaller treatment groups and thus for any combination of imbalance in treatment and control arm sample sizes across the studies. Given the observed data, the possible outcomes for the $i$th study under are a balanced design can be enumerated as
\begin{align*}
    \Dsc_i^* = \{ (Y_{i1l}^*, N_{i2}), (Y_{i2}, N_{i2}) \mid  l = \max(0, N_{i2}-N_{i1}+Y_{i1}), \dots, Y_{i1} \}
\end{align*}
where $Y_{i1l}^* = l$.  The probability of observing $l$ events in the treated arm is dictated by the hypergeometric distribution as 
$$ p_{i1l}  = \frac{ { N_{i1} - Y_{i1}  \choose N_{i2} - l }   {Y_{i1}  \choose  l } }{  {N_{i1} \choose N_{i2} }}.$$
For example, suppose the data for two studies is 
\begin{align*}
\Dsc_1^O &= \{ (Y_{11}, N_{11}),  (Y_{12}, N_{12}) \} =  \{ (2, 80), (1, 50) \} \\
\Dsc_2^O &= \{ (Y_{21}, N_{21}),  (Y_{22}, N_{22}) \} =  \{ (1, 100), (0, 90) \}
\end{align*}
For the first study, the possible outcomes under the balanced design are given by
\begin{align*}
\Dsc_1^* &= \{ (Y_{11l}^*, N_{12}),  (Y_{12}, N_{12}) \mid l = 0, \dots Y_{11} \} \\
&= \{ [(0, 50), (1, 50)], [(1, 50), (1, 50)], [(2, 50), (1, 50)] \}.
\end{align*}
Here we have sampled 50 of the 80 patients in first study's treatment arm, 50 being the size of the smaller arm, of which up to 2 experience events. 

For the second study, we similarly obtain $\Dsc_2^* =\{ [(0, 90), (0, 90)], [(1, 90), (0, 90)] \}$.  In this study, and more generally when there is a study with zero events in the control arm, we obtain a DZ study.  As our approach removes DZ studies, we employ the following correction to proceed with moment estimation as in Section \ref{sec: balan}:
$${\Dsc}_i^{*c} = \{ (Y_{i1l}^*, N_{i1l}),  (Y_{i2}, N_{i2}) \mid l \in \Lsc \} \quad \mbox{where} \quad \Lsc  = \begin{cases} 0, \dots, Y_{i1} \quad \mbox{if} \quad Y_{i2} \ne 0 \\ 
1, \dots, Y_{i1} \quad \mbox{if}  \quad Y_{i2} = 0  \end{cases} $$
and $p_{i1l}^c =  p_{i1l}/\sum_{l \in \Lsc} p_{i1l}$.  Letting 
$Y_{i\cdot l}^* = Y_{i1l}^* + Y_{i2}$, we then estimate $\mu_0$ and $\nu_0$ based on $\Dsc_i^{*c}$ and $p_{i1l}^c$  as
\begin{equation*}
\mutilde = K^{-1} \sum_{i = 1}^K \sum_{l \in \Lsc} \frac{Y_{i1l}^*}{Y_{i\cdot l}^* } p_{i1l}^c \quad \mbox{and} \quad \nutilde =  \max \left\{ 0,   \frac{ \sum_{i = 1}^K  \sum_{l \in \Lsc} \left\{ \left( \frac{\Ytilde_{i1l}^*} {\Ytilde_{i\cdot l}^*} \right)^2 -   \frac{\mutilde_{int} }{\Ytilde_{i\cdot l^*}} \right\} p_{i1l}^c  }{\sum_{i = 1}^K  \sum_{l \in \Lsc} \left( 1 - \frac{1} {\Ytilde_{i\cdot l}^*} \right) p_{i1l}^c }- \mutilde_{int}^2 \right\}
\end{equation*}
where $\mutilde_{int} = K^{-1} \sum_{i = 1}^K  \sum_{l \in \Lsc}\frac{\Ytilde_{i1l}^*} {\Ytilde_{i\cdot l}^*}p_{i1l}^c $ and $(\Ytilde_{i1l}^*, \Ytilde_{i\cdot l}^*) = (Y_{i1l}^*+ 0.5, Y_{i\cdot l}^*+ 1)$. The test statistic is taken as  
\begin{equation*}
T(\mu; \Dsc^0) = \frac{(\mutilde - \mu)^2}{ \widehat{\text{Var}}(\mutilde )} \quad \mbox{where} \quad  \widehat{\text{Var}}(\mutilde )=K^{-2} \left\{ \sum_{i = 1}^K \sum_{l \in \Lsc} p_{i1l}^c \frac{\mutilde(1-\mutilde)}{\Ytilde_{i\cdot l}^*} + p_{i1l}^c\left(1- \frac{1}{\Ytilde_{i\cdot l}^*}\right)\nutilde   \right\}.
\end{equation*}

Though not straightforward to verify analytically, our numerical studies coincide with the intuition that our proposed estimator is consistent for the first and second moments in settings with sufficiently high event rates in the control arm.  Our adjustment for DZ studies biases the point estimator away from the null with the magnitude of bias depending on the degree of imbalance and the rarity of the event.  The validity of XRRmeta still holds, however, as inference is based on the exact distribution of $T(\mu; \Dsc^0)$. The bias only affects efficiency and therefore the width of the resulting confidence interval. Our numerical studies indicate that our proposed test statistic does not result in overly conservative inference suggesting that efficiency is not greatly affected.

 \subsection{Computational Details}
Another useful feature of our proposed test statistic is that $T\left(\mu; \Dsc^{\mu, \nu_1}\right)$ appears to generally first order stochastically dominate $T\left(\mu; \Dsc^{\mu, \nu_2}\right)$ for a fixed $\mu$ and $\nu_1 < \nu_2$.  That is, our numerical studies suggest that 
\begin{equation*}
P\left\{ T\left(\mu; \Dsc^{\mu, \nu_1}\right) > t\right\}  < P\left\{ T\left(\mu; \Dsc^{\mu, \nu_2}\right) > t\right\} 
\end{equation*}
for $t$ in the tail region of interest.  Intuitively, this property follows from the fact that a random effects distribution with higher variability will yield a test statistic with more variation and hence a larger tail probability \supercite{gronsbell2018}.  Practically, this property may be leveraged to significantly decrease execution time as it implies  
$$p(\mu; \Dsc^0)=\sup_v p(\mu, v; \Dsc^0)\approx p\left\{\mu, \nu_{\sup}(\mu); \Dsc^0\right\} $$
Computational complexity can be reduced by $O(J)$ operations by computing $p(\mu; \Dsc^0)$ with the values along the boundary of the reduced parameter space. 

Supplementary Figure S1 details the complete execution of XRRmeta.  The procedure involves three steps: Initialization, Iteration, and Correction.  The initialization step is used to further accelerate computation by beginning the grid search using asymptotic confidence bounds whenever possible.  The iteration step leverages the stochastic dominance result to move along the boundary of the restricted parameter space to determine initial upper and lower limits of the CI.  The correction step checks values of $(\mu, \nu)$ beyond the limits found in the iteration step as the stochastic dominance result is only an approximation.  In terms of the implementation of XRRmeta, a key decision is the choice of grid size, $s$, used to iterate along the boundary of the parameter space.  We suggest using the original scale to explicitly control the precision of the calculations.  Taking $s = 0.001$ is reasonable for most effect sizes and was utilized in our simulation and real data analyses. The number of Monte Carlo iterations similarly depends on the desired precision. For example, if we aim to evaluate a $p$ value of 0.05 with a standard error of 0.005 then we must take $M \ge 2,000$. The number of $(\mu, \nu)$ pairs to evaluate in the correction step can simply be taken as a multiple of $s$.  We found that 10$s$ was sufficient in our numerical studies.

\section{Simulation Studies}\label{sec: sim}
We evaluated the performance of XRRmeta through extensive simulation studies.  In all settings, the observed data were generated as
\begin{align*}
   Y_{ij} \sim \text{Poisson}( N_{ij} \lambda_{ij} ), \text{ }\lambda_{i1} \sim  \text{Gamma}\left(\alpha_0 , 
   \alpha_0 / r_0 \right), \text{ and } \lambda_{i2} \sim  \text{Gamma}\left(\beta_0 ,  \alpha_0 / r_0 \right)
\end{align*}
to achieve an average event rate of $r_0$ in the treated arm and an average event rate of $r_0\beta_0/\alpha_0$ in the control arm.  As the scale parameters of the gamma distribution for the both arms are identical, $\pi_i \sim \text{Beta}(\alpha_0, \beta_0)$.  We varied (i) the total number of studies, $K_{tot}$, (ii) the event rate in the treated arm, $r_0$, and (iii) the values of $\alpha_0$ and $\beta_0$ to represent varying degrees of treatment effect and between-study heterogeneity.  The three primary settings, representing high, moderate, and low between-study heterogeneity, are summarized in Table \ref{tab: sim}. \\

\begin{table}[H]
 \caption{Parameters of the Beta distribution for the three primary simulation settings.}
 \label{tab: sim}
\begin{center}
\begin{tabular}{| l  | l | l | }
  \hline
{\bf{Setting}} & {\bf{Treatment Effect}} & $(\alpha_0, \beta_0)$ \\ 
\hline
\hline
1: High Heterogeneity & Null & (1.45, 1.45)  \\
 & Protective & (1.10, 1.65)    \\
 \hline
 2: Moderate Heterogeneity & Null & (5.50, 5.50)   \\
 & Protective & (4.20, 6.30)  \\
\hline
3: Low Heterogeneity & Null & (145, 145)   \\
 & Protective & (110, 165)  \\
 \hline
  \end{tabular}
\end{center}
\end{table}

In the settings with a protective effect, the values of $(\alpha_0, \beta_0)$ were selected to achieve a relative risk of $0.67$  ($\mu_0 = 0.4$) and to maintain a similar between-study heterogeneity to the setting of no treatment effect.  The between study variation for Settings 1, 2, and 3 is approximately $ \nu_0 = 0.064$, $0.021$, and $0.001$, respectively.  In all settings, we considered $r_0 = 0.01$  and 0.03 with the number of studies $K = 12$, $24$, $48$, and $96$. The percentage of DZ studies was approximately 15\% on average across the settings with $r_0 = 0.01$ and near 0\% with $r_0 = 0.03$.  The study specific sample sizes $(N_{i1}, N_{i2})$ were randomly sampled from the rosiglitazone study reported in Supplementary Table S1.  All results were averaged over 2000 replications.   

To assess the conservativeness and efficiency of XRRmeta while facilitating comparisons across commonly used existing meta-analytic methods for the odds ratio, we summarized the type I error for the null effect settings and the power for the protective effect settings.  As the odds ratio and relative risk are comparable in the rare events setting we only report results for the former.  We compared XRRmeta to five common methods, including the Mantel-Haenszel method with and without a 0.5 continuity correction for zero event studies (MH, MH-CC), the fixed and random effects Peto method (Peto-F, Peto-R), and the DerSimonian-Laird method with a 0.5 continuity correction for zero event studies (DL).  These methods were implemented with the {\it{metabin}} package in {\texttt{R}}.

The results for all three settings with $r_0 = 0.01$ are summarized in Figure \ref{fig: sim_result}.  Methods with inflated type I error are presented in a lighter shade as the power may not be properly interpreted \supercite{christensen1997comparison}.  Overall, XRRmeta is the only method that consistently controls type I error across sample sizes and heterogeneity levels.  The setting of high between-study heterogeneity is particularly striking as none of the comparators provides a valid test of the treatment difference at any of the study sizes considered.  Moreover, all of the comparators with the exception of the random effects Peto method exhibit the counterintuitive property that the type I error control deteriorates as the number of studies meta-analyzed increases, precisely the setting where practitioners might expect the conclusions to become more reliable. 

The fixed effect approaches (Mantel-Haenszel with or without continuity correction and Peto fixed effect) have substantially inflated type I error in the high and moderate heterogeneity settings. The random effects comparators (Peto random effects and DerSimonian-Laird) better cope with between-study heterogeneity, but still do not maintain the type I error. With an increasing number of studies, the type I error of the Peto random effects method approaches the nominal level. Yet even with $K = 96$ (more than are typically available for meta-analysis), the Peto random effects method did not provide proper control outside of the low heterogeneity setting.  Similar to the fixed effect approaches, the DL method exhibited type I error increasing with the number of studies, a phenomenon that has been reported previously \supercite{cai2010meta}. Although increasingly conservative as the between-study heterogeneity declines and less powerful than the alternative random effects approaches, XRRmeta is the only method that robustly controls the type I error.  Lastly, we note that we also examined the empirical coverage level of the constructed confidence intervals across different settings. Note that one minus the type I errors reported in Figure \ref{fig: sim_result} are the observed coverage levels of confidence intervals under the null. The empirical coverage levels are always above 95\% as the proposed exact procedures ensured.  In addition, the coverage levels in most settings are below 98\%, suggesting that the resulting exact intervals are not overly conservative.

\begin{figure}[H]
\begin{subfigure}[b]{\textwidth}
\centering
\caption{Setting 1: High Heterogeneity}
\includegraphics[scale = 0.13]{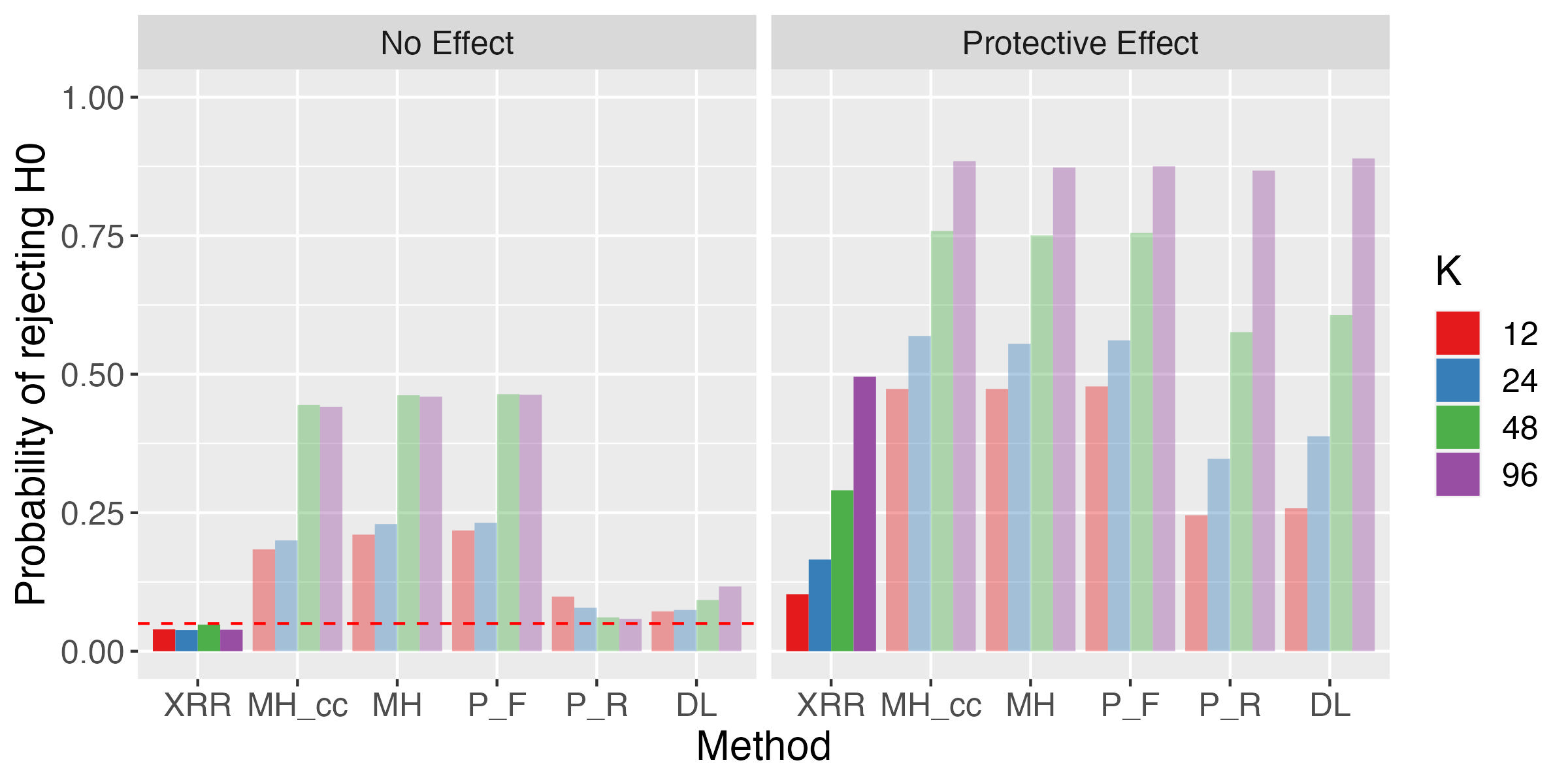}
\end{subfigure}
\begin{subfigure}[b]{\textwidth}
\centering
\caption{Setting 2: Moderate Heterogeneity}
\includegraphics[scale = 0.13]{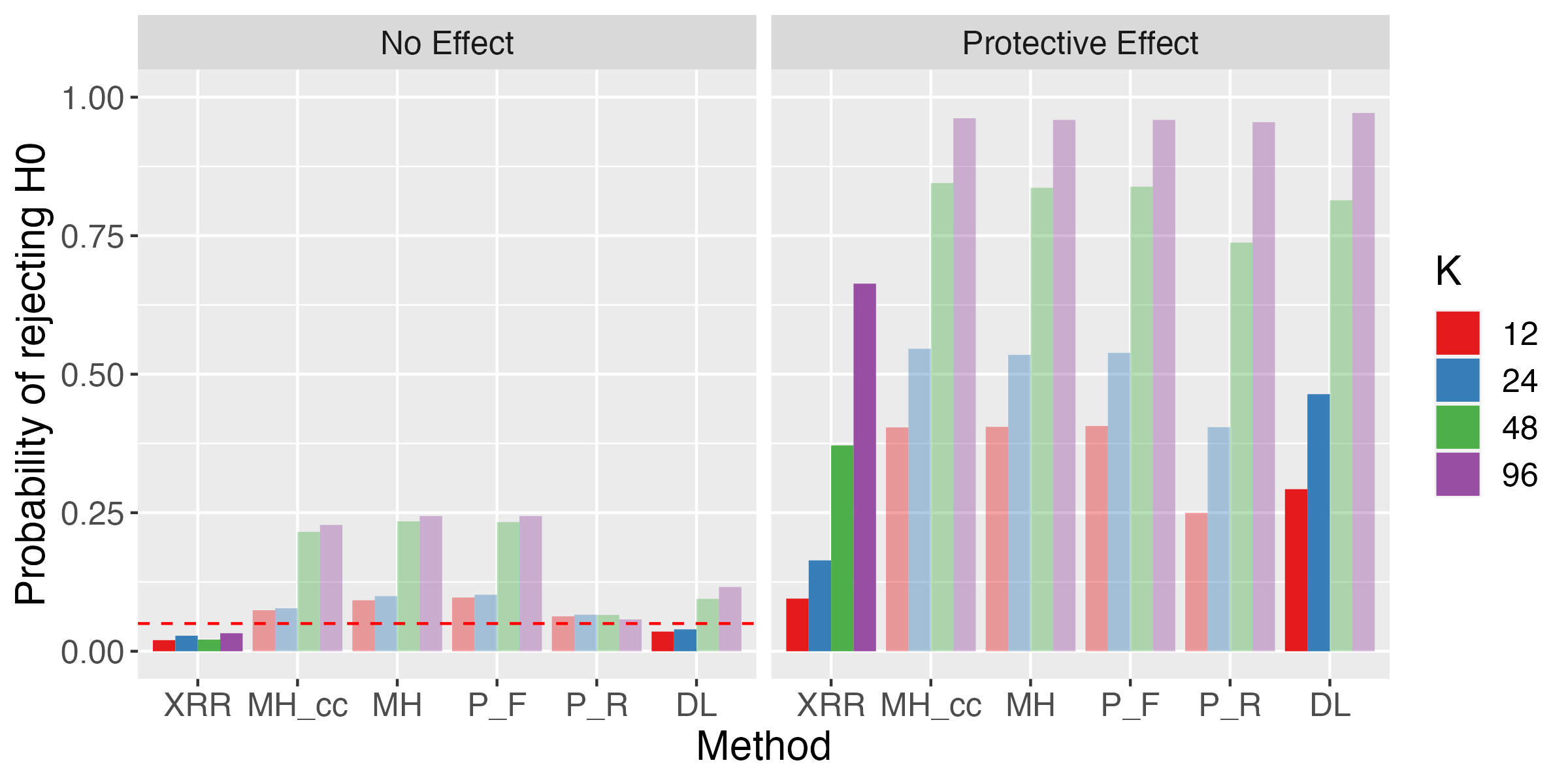}
\end{subfigure}
\begin{subfigure}[b]{\textwidth}
\centering
\caption{Setting 3: Low Heterogeneity}
\includegraphics[scale = 0.13]{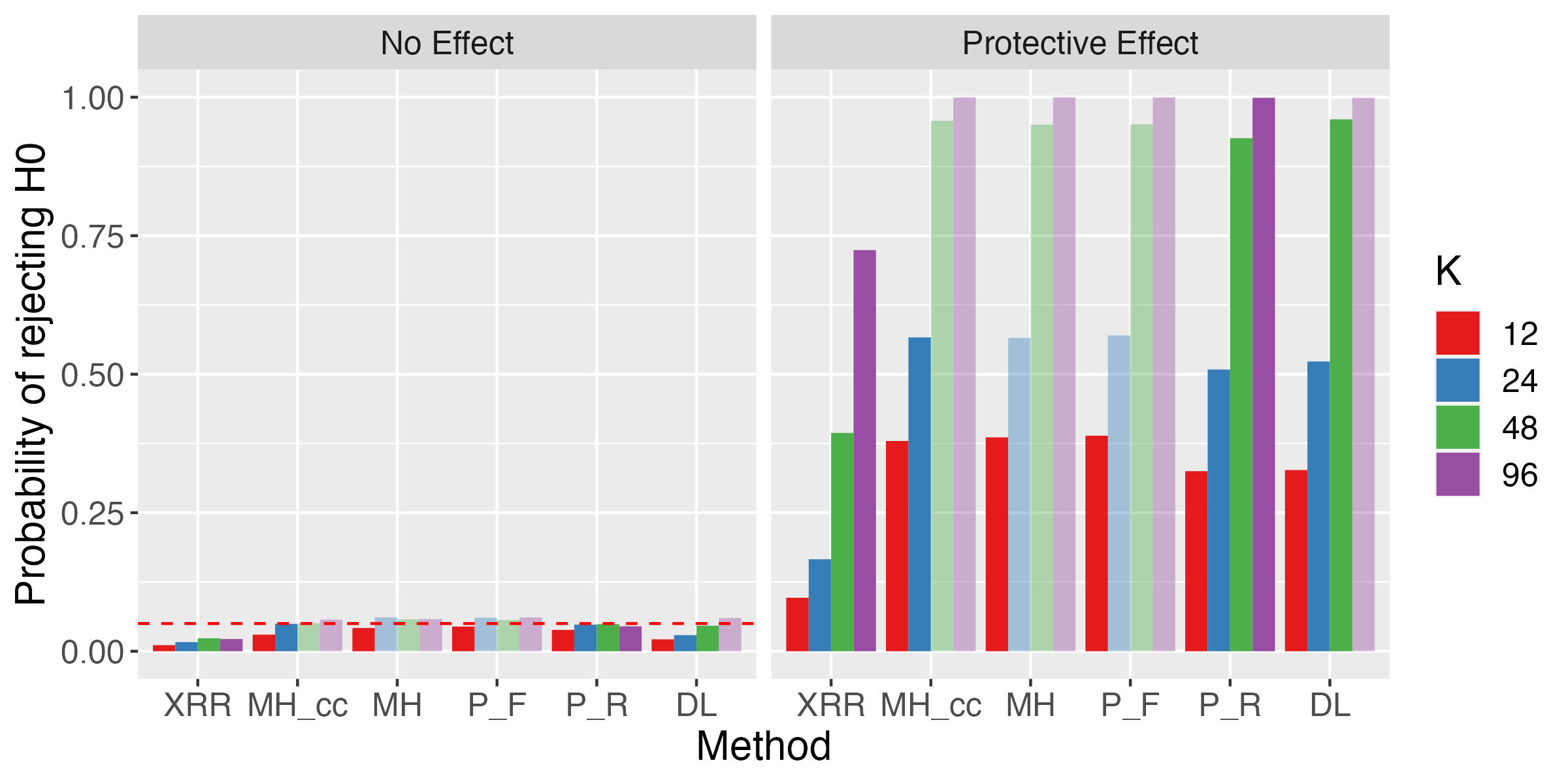}
\end{subfigure}
\caption{Type I error and power with $r_0 = 0.01$ for XRRmeta (XRR), Mantel-Haenszel with and without a 0.5 continuity correction (MH, MH\_cc), the fixed and random effects Peto method (P\_F, P\_R), and the DerSimonian-Laird method with a 0.5 continuity correction (DL).  Methods that do not control the type I error are shown in a lighter shade.}\label{fig: sim_result}
\end{figure}

The results for $r_0 = 0.03$ are presented in Supplementary Figure S1.  The same patterns are observed, but higher power is achieved for all methods due to the higher event in the treated group.  Together, our results demonstrate the utility of fixed effect methods when between-study homogeneity is expected and the benefit of XRRmeta in uniformly controlling type I error when between-study heterogeneity is expected.

\section{Real Data Examples}\label{sec: real}
We next evaluate the performance of the proposed procedure in two real data examples.  The first is our motivating example, the rosiglitazone study, where the proportion of DZ studies is high as the events of interest are extremely rare.  The second involves a recent meta-analysis of the utility of face masks in preventing person-to-person transmission of SARS-CoV-2 and COVID-19, which has fewer studies, but a relatively higher event rate compared to the rosiglitazone study.  For both analyses, we present the results from XRRmeta and the 5 comparison methods utilized in the simulation studies (MH, MH-CC, Peto-F, Peto-R, DL). 

\subsection{Rosiglitazone Study} 
Rosiglitazone is a popular drug for the treatment of type II diabetes mellitus.  Its effect on cardiovascular mortality (CVD) and myocardial infarction (MI) has been under scrutiny since the 2007 publication of Niessen and Wolski's meta-analysis of 42 randomized controlled clinical trials in the {\it{New England Journal of Medicine}}\supercite{nissen2007effect}.  The data depicted in Supplementary Table S1 includes the 42 studies reported in the original paper as well as 6 additional DZ studies that were excluded from analysis.  In the original analysis, the authors utilized the fixed-effect Peto method. They found that in the rosiglitazone group, as compared with the control group, the odds ratio for MI was 1.43 (95\% CI: [1.03, 1.98]; $p$ = 0.03) while the odds ratio for CVD was 1.64 (95\% CI: [0.98, 2.74]; $p$ = 0.06). However, these conclusions are questionable for numerous reasons, including the exclusion of DZ trials and the use of a fixed-effect approach.  For instance, including the DZ studies and employing the Mantel-Haenszel method with a continuity correction of 0.5 in all studies results in a contradictory conclusion for the MI endpoint (Table \ref{tab: rog comp}).  \cite{tian2008exact} obtained similar results in their re-analysis of the data using their proposed exact confidence procedure for the risk difference under a fixed-effect model.  It is also worth noting that a conventional fixed-effect approach is not appropriate for the rosiglitazone data as two very large studies are present in the data analysis \supercite{shuster2007fixed}.  For example, excluding the two larger studies from Niessen and Wolski's analysis yields substantially different intervals of [0.88, 2.39] and [1.17, 4.91] for MI and CVD, respectively. A random effects approach that weighs all studies equally is more appropriate for the rosiglitazone data.     

We report the results for the treatment contrast from XRRmeta and the comparisons methods for the odds ratio in Table \ref{tab: rog comp}.  In terms of the comparison methods, the CIs from the fixed effect Peto method correspond to the original analysis for both outcomes because the 6 additional DZ studies do not influence the results.  For the MI outcome, the random effects Peto CI is identical to the fixed effect model.  Both of the Peto CIs and the Mantel-Haenszel CI without continuity correction coincide with the original conclusions of a significant effect. Adding a continuity correction to the Mantel-Haenszel analysis or utilizing the Dersimonian-Laird method reverses the conclusion. For CVD, these two methods demonstrate that there is no statistically significant evidence of increased risk of CVD associated with rosiglitazone while the  Mantel-Haenszel method without continuity correction and both Peto methods show marginal significance of a positive effect.

\begin{table}[H]
\caption{Meta-analysis results for the myocardial infarction (MI) and cardiovascular death (CVD) endpoints of the rosiglitazone study. Point estimates of the odds ratios, the 95\% confidence intervals (CI), and p-values from the Mantel-Haenszel method with and without a 0.5 continuity correction (MH, MH-CC), the fixed and random effects Peto method (Peto-F, Peto-R), and the DerSimonian-Laird method with a 0.5 continuity correction for zero event studies (DL) as well as the treatment contrast results from XRRmeta.  Statistically significant results are in bold.}\label{tab: rog comp}
\centering
\begin{tabular}{lllllll|}
   \hline
Endpoint &  Method & Point Estimates & CI & CI length & $p$-value  \\ 
\hline
  \hline
MI & MH  & 1.42 &  {{[1.03, 1.98]}} & 0.95 &  {\bf{0.033 }}  \\ 
& MH-CC  &  1.23 & [0.92, 1.65] & 0.73 & 0.163    \\ 
& Peto-F  & 1.43   & {{[1.03, 1.98]}} & 0.95 & {\bf{0.032 }} \\ 
& Peto-R  & 1.43   & {{[1.03, 1.98]}} & 0.95 &  {\bf{0.032}}  \\ 
& DL  & 1.23 &  [0.91, 1.67] & 0.76 &  0.178  \\ 
\hline
& XRRmeta & 0.67  & {{[0.51, 0.82]}} & 0.31 &  {\bf{0.047}} \\ 
   \hline
&  &   &  &  &  \\

CVD & MH  & 1.70 &  {{[0.98, 2.93]}} & 1.95 & 0.057  \\ 
& MH-CC  & 1.13 &  {{[0.76, 1.69]}}  & 0.93  & 0.541  \\ 
& Peto-F  &  1.64  &  {[0.98, 2.74]} & 1.76 &  0.060 \\ 
& Peto-R &  1.64   &  {[0.98, 2.74]} & 1.76 &  0.060  \\ 
& DL  &  1.10 &  {[0.73, 1.66]} & 0.93 &  0.662  \\ 
\hline
& XRRmeta &  0.79  & {\bf{[0.56, 0.90]}} & 0.34 &  {\bf{0.010}}  \\ 
   \hline\\
\end{tabular}
\end{table}

In the XRRmeta analysis, both exact confidence intervals exclude the null value of 0.5.  More specifically, our results are consistent with the original study for the MI endpoint, indicating that there is some evidence for an increased risk of MI among patients receiving rosiglitazone.  However, we obtain conflicting results for CVD.  Our analysis suggests the frequency of CVD in the treated arm is higher than the control arm, which is consistent with the random effects analysis of Shuster et al.\ \supercite{shuster2007fixed}.  Unlike the 5 comparisons methods, we reiterate that our procedure is justified in its exclusion of DZ studies and is guaranteed to provide valid inference in settings with a relatively small number of studies and between-study variability. This difference in conclusion could be due to the large number of DZ studies, presence of the two larger studies, and/or between study heterogeneity in the CVD analysis.

\subsection{Face Mask Study} 
COVID-19 is caused by severe acute respiratory syndrome coronavirus 2 (SARS-CoV-2) and is transmitted from person-to-person through close contact.  In the midst of the COVID-19 pandemic,  \cite{chu2020physical} conducted a systematic review to evaluate the preventative effect of face masks on virus transmission in healthcare and community settings.  Twenty-nine studies met the inclusion criteria for their study and the resulting data are presented in Supplementary Table S2.  The authors obtained an estimated relative risk of 0.34 (95\% CI: [0.26, 0.45], $p < 0.0001$) with the DL method excluding the DZ studies, indicating substantially reduced transmission with face masks. 

We report the results for the treatment contrast from XRRmeta and the comparisons methods for the odds ratio in Table \ref{tab: face mask comp}.  The CIs from all 5 comparison methods yield similar results, with the random effects intervals having longer length as expected.  In contrast to the original analysis, including the DZ studies provides stronger evidence of a protective effect of face masks.  XRRmeta yields a conclusion consistent with the results of the comparison methods as the exact confidence interval for $\mu_0$ is well below 0.5.  Overall, there is clear evidence of reduced person-to-person virus transmission with face mask use across all our analyses.  Moreover, these results echo our simulation studies which illustrate that XRRmeta is not substantially underpowered relative to its classical counterparts.  \\

\begin{table}[H]
\caption{Meta-analysis results for the face mask study. Point estimates of the odds ratios, the 95\% confidence intervals (CI), and p-values from the Mantel-Haenszel method with and without a 0.5 continuity correction (MH, MH-CC), the fixed and random effects Peto method (Peto-F, Peto-R), and the DerSimonian-Laird method with a 0.5 continuity correction for zero event studies (DL) as well as the treatment contrast results from XRRmeta. }\label{tab: face mask comp}
\centering
\begin{tabular}{lllllll|}
   \hline
  Method & Point Estimates & CI & CI length & $p$-value  \\ 
\hline
  \hline
MH  &  0.22 & {{[0.18, 0.28]}}  & 0.10 & $< 0.0001$   \\ 
 MH-CC  & 0.23  & {{[0.18, 0.28]}}   & 0.10 & $< 0.0001$   \\ 
 Peto-F  & 0.27 & {{[0.22, 0.32]}} &  0.10 & $< 0.0001$\\ 
 Peto-R  & 0.24 & {{[0.18, 0.33]}}  & 0.15   & $< 0.0001$\\ 
 DL  & 0.22  & {{[0.16, 0.32]}} &  0.16 & $< 0.0001$ \\ 
\hline
XRRmeta &  0.19   & {{[0.11, 0.27]}} & 0.16 &  $< 0.005$\\ 
   \hline\\
\end{tabular}
\end{table}

 \section{Discussion}\label{sec: discussion}
We introduced a new method, XRRmeta, for performing a random effects meta-analysis of the treatment effect in a two-group comparison. Unlike classical methods, the coverage of the confidence interval from our method is guaranteed to be at or above the nominal level (up to Monte Carlo error) in settings with rare events, high between-study heterogeneity, and few or small studies. XRRmeta is also justified in its exclusion of zero-event studies through a conditional inference argument.  As noted by Zabriskie et al., the current setting has been largely underappreciated in the meta-analysis literature.  The {\it{Cochrane Handbook for Systematic Reviews of Interventions}} suggests that ``incorporation of heterogeneity into an estimate of a treatment effect should be a secondary consideration when attempting to produce estimates of effects from sparse data - the primary concern is to discern whether there is any signal of an effect in the data" \supercite{zabriskie2021permutation, higgins2011cochrane}.  Our numerical studies, however, illustrate that the presence of heterogeneity has a large bearing on the conclusions drawn from meta-analyses in such settings.  XRRmeta was the only method that uniformly maintained type I error while not yielding overly conservative inference. The utility of XRRmeta was also demonstrated through analyses of the rosiglitazone study and a study of the protective effects of face masks \supercite{nissen2007effect, chu2020physical}.  Moreover, we have released an \texttt{R} package implementing XRRmeta to encourage use by practitioners (\href{https://github.com/zrmacc/RareEventsMeta}{https://github.com/zrmacc/RareEventsMeta}).   
 
It is important to note that the performance of our procedure inherently depends on the choice of test statistic.  Our numerical studies and real data analyses indicate that the chosen Wald test statistic provides a reasonable balance between statistical efficiency and computational speed.  For example, the rosiglitazone study took roughly 30 minutes to run (Intel Core i7-1060NG7 $@$ 1.2 GHz).  Additionally, we did not explicitly consider estimation of between-study heterogeneity, which is challenging in the small $K$ and/or rare event setting.  Lastly, knowledge of potential differences across studies can guide practitioners in the choice of a fixed effect or random effects approach.  We also highlight that key consideration in the use of XRRmeta relative to classical counterparts is the price paid in terms of power to control type I error.  While XRR method favors conservatism, it is ultimately a question of the problem at hand if preserving type I error is the priority.



 \section{Acknowledgement}\label{sec: Acknowledgement}
J.G. is supported by an NSERC Discovery Grant (RGPIN-2021-03734), a University of Toronto Data Science Institute Methodology Seed Grant, and a Connaught New Researcher Award. L.T. is supported by an RO1 grant 5R01HL08977812 from National Institutes of Health. 


\begin{singlespace}
\printbibliography
\end{singlespace}

\end{document}


\makeatletter
\renewcommand{\fnum@figure}{\figurename~S\thefigure}
\makeatother

\makeatletter
\renewcommand{\fnum@table}{\tablename~S\thetable}
\makeatother

\maketitle

\section{Rosiglitazone Data}
\begin{table}[H]
\centering
\scalebox{0.72}{\begin{tabular}{lllllll}
  \hline
  \hline
  & \multicolumn{3}{c}{Rosiglitazone} & \multicolumn{3}{c}{Control} \\
\hline
Study ID & N & CVD & MI & N & CVD & MI \\ 
\hline
  1 & 357 & 1 & 2 & 176 & 0 & 0 \\ 
  2 & 391 & 0 & 2 & 207 & 0 & 1 \\ 
  3 & 774 & 0 & 1 & 185 & 0 & 1 \\ 
  4 & 213 & 0 & 0 & 109 & 0 & 1 \\ 
  5 & 232 & 1 & 1 & 116 & 0 & 0 \\ 
  6 & 43 & 0 & 0 & 47 & 0 & 1 \\ 
  7 & 121 & 0 & 1 & 124 & 0 & 0 \\ 
  8 & 110 & 3 & 5 & 114 & 2 & 2 \\ 
  9 & 382 & 0 & 1 & 384 & 0 & 0 \\ 
  10 & 284 & 0 & 1 & 135 & 0 & 0 \\ 
  11 & 294 & 2 & 0 & 302 & 1 & 1 \\ 
  12 & 563 & 0 & 2 & 142 & 0 & 0 \\ 
  13 & 278 & 0 & 2 & 279 & 1 & 1 \\ 
  14 & 418 & 0 & 2 & 212 & 0 & 0 \\ 
  15 & 395 & 2 & 2 & 198 & 0 & 1 \\ 
  16 & 203 & 1 & 1 & 106 & 1 & 1 \\ 
  17 & 104 & 0 & 1 & 99 & 0 & 2 \\ 
  18 & 212 & 1 & 2 & 107 & 0 & 0 \\ 
  19 & 138 & 1 & 3 & 139 & 0 & 1 \\ 
  20 & 196 & 1 & 0 & 96 & 0 & 0 \\ 
  21 & 122 & 0 & 0 & 120 & 0 & 1 \\ 
  22 & 175 & 0 & 0 & 173 & 0 & 1 \\ 
  23 & 56 & 0 & 1 & 58 & 0 & 0 \\ 
  24 & 39 & 0 & 1 & 38 & 0 & 0 \\ 
  25 & 561 & 1 & 0 & 276 & 0 & 2 \\ 
  26 & 116 & 2 & 2 & 111 & 1 & 3 \\ 
  27 & 148 & 2 & 1 & 143 & 0 & 0 \\ 
  28 & 231 & 1 & 1 & 242 & 0 & 0 \\ 
  29 & 89 & 0 & 1 & 88 & 0 & 0 \\ 
  30 & 168 & 1 & 1 & 172 & 0 & 0 \\ 
  31 & 116 & 0 & 0 & 61 & 0 & 0 \\ 
  32 & 1172 & 1 & 1 & 377 & 0 & 0 \\ 
  33 & 706 & 1 & 0 & 325 & 0 & 0 \\ 
  34 & 204 & 0 & 1 & 185 & 1 & 2 \\ 
  35 & 288 & 1 & 1 & 280 & 0 & 0 \\ 
  36 & 254 & 0 & 1 & 272 & 0 & 0 \\ 
  37 & 314 & 0 & 1 & 154 & 0 & 0 \\ 
  38 & 162 & 0 & 0 & 160 & 0 & 0 \\ 
  39 & 442 & 1 & 1 & 112 & 0 & 0 \\ 
  40 & 394 & 1 & 1 & 124 & 0 & 0 \\ 
  41 & 2635 & 12 & 15 & 2634 & 10 & 9 \\ 
  42 & 1456 & 2 & 27 & 2895 & 5 & 41 \\ 
  43 & 101 & 0 & 0 & 51 & 0 & 0 \\ 
  44 & 232 & 0 & 0 & 115 & 0 & 0 \\ 
  45 & 70 & 0 & 0 & 75 & 0 & 0 \\ 
  46 & 25 & 0 & 0 & 24 & 0 & 0 \\ 
  47 & 196 & 0 & 0 & 195 & 0 & 0 \\ 
  48 & 676 & 0 & 0 & 225 & 0 & 0 \\ 
   \hline
\end{tabular}}
\caption{Data for the rosiglitazone study. Shown are the study sizes (N), number of myocardial infarctions (MI), and number of cardiovascular deaths (CVD) for the treated and control arms.}\label{tab: rog data}
\end{table}

\section{Computational Details of XRRmeta }
Here we detail the three steps involved in implementing XRRmeta: Initialization, Iteration, and Correction. Let $s$ denote the step size for the grid along the $\mu$ axis and $k$ a positive integer.
\begin{description}
\item[Initialization Step.] Obtain starting points for the iteration step.
\begin{description}
\item[Step a.] Compute $(\mu^{MOM}_{LB}, \mu^{MOM}_{UB})$, the CI based on the asymptotic $\chi^2$ approximation to $\mutilde$.
\item[Step b.] Evaluate 
$$\tilde{p}(\mu^{\widetilde{MOM}}_{LB}; \Dsc^0) \text{ and } \tilde{p}(\mu^{\widetilde{MOM}}_{UB}; \Dsc^0)$$
where $\tilde{p}(\mu; \Dsc^0)=p\left\{\mu, \nu_{\sup}(\mu); \Dsc^0\right\}$, 
$\mu^{\widetilde{MOM}}_{LB} = \max(s, \mu^{MOM}_{LB})$, and $\mu^{\widetilde{MOM}}_{UB} = \min(1-s, \mu^{MOM}_{UB})$.
 \item[Step c.] Take the upper and lower starting values as 
$$\mutilde_{\text{inf}} =\mu^{\widetilde{MOM}}_{LB} I \left\{ \tilde{p}(\mu^{\widetilde{MOM}}_{LB}; \Dsc^0) \ge \alpha \right\} + \mutilde  I \left\{ \tilde{p}(\mu^{\widetilde{MOM}}_{LB}; \Dsc^0) < \alpha \right\}$$ 
and
$$\mutilde_{\text{sup}} = \mu^{\widetilde{MOM}}_{UB} I \left\{ \tilde{p}(\mu^{\widetilde{MOM}}_{UB}; \Dsc^0) \ge \alpha \right\} + \mutilde  I \left\{ \tilde{p}(\mu^{\widetilde{MOM}}_{UB}; \Dsc^0) < \alpha \right \}.$$    
\end{description}
\item[Iteration Step.] Iterate along the grid on the $\mu$ axis based on the initialization points.
\begin{description}
\item[Step a.] Find the upper bound by iterating out from $\mutilde_{\text{sup}}$ until 
$$\mu^{\text{it}}_{UB} = \inf_{\mu} \left[ \mu \mid \tilde{p}(\mu; \Dsc^0) \ge \alpha \right].$$
\item [Step b.] Find the lower bound by iterating out from $\mutilde_{\text{inf}}$ until 
$$\mu^{\text{it}}_{LB} = \sup_{\mu} \left[ \mu \mid \tilde{p}(\mu; \Dsc^0)\ge \alpha \right].$$
\end{description}
\item[Correction Step.] Evaluate a $(\mu, \nu)$ grid beyond the bounds from the iteration step.
\begin{description}
\item[Step a.] Obtain $p$ values for $\mu$ in $(\mu^{\text{it}}_{UB} , \mu^{\text{it}}_{UB} + ks)$ and $(\mu^{\text{it}}_{LB} - ks, \mu^{\text{it}}_{LB})$ and let
$$\mu_{UB} = \max\{ \mu^{\text{it}}_{UB} + ks \mid  \tilde{p}(\mu^{\text{it}}_{UB} + ks; \Dsc^0)\ge \alpha, k \ge 0  \} \text{ and }$$
$$\mu_{LB} = \min\{ \mu^{\text{it}}_{LB} - ks \mid  \tilde{p}( \mu^{\text{it}}_{LB} - ks; \Dsc^0)\ge \alpha, k \ge 0  \}. $$ 
\item [Step b.] Compute $p(\mu; \Dsc^0) = \sup_{\nu} p(\mu, \nu; \Dsc^0)$ for $\mu \in (\mu_{UB}, \mu_{UB} + \delta)$ and  $\mu \in (\mu_{LB} - \delta, \mu_{LB})$ for some small $\delta >0$.  For example, for the upper bound, compute $p$ values along $ j = 1, \dots, J$ equally spaced values of $\mu^{j} \in (\mu_{UB}, \mu_{UB} + \delta)$ and corresponding $ i = 1, \dots, I_j$ equally spaced values of $\nu^{i_j} \in (0, \nu_{\sup}(\mu))$ to approximate $p(\mu^j; \Dsc^0)$ as
$$\max \{ p(\mu^j, \nu^{i_j}; \Dsc^0) \mid i = 1, \dots, I_j  \}.$$
\item [Step c.] Compute the upper bound of the interval as 
$$ \max\{\mu^j \mid p(\mu^j; \Dsc^0) \ge \alpha, j = 1, \dots, J  \} $$
and the lower bound in a similar manner.
\end{description}
\end{description}

\section{Additional Simulation Results}
\begin{figure}[H]
\begin{subfigure}[b]{\textwidth}
\centering
\caption{Setting 1: High Heterogeneity}
\includegraphics[scale = 0.12]{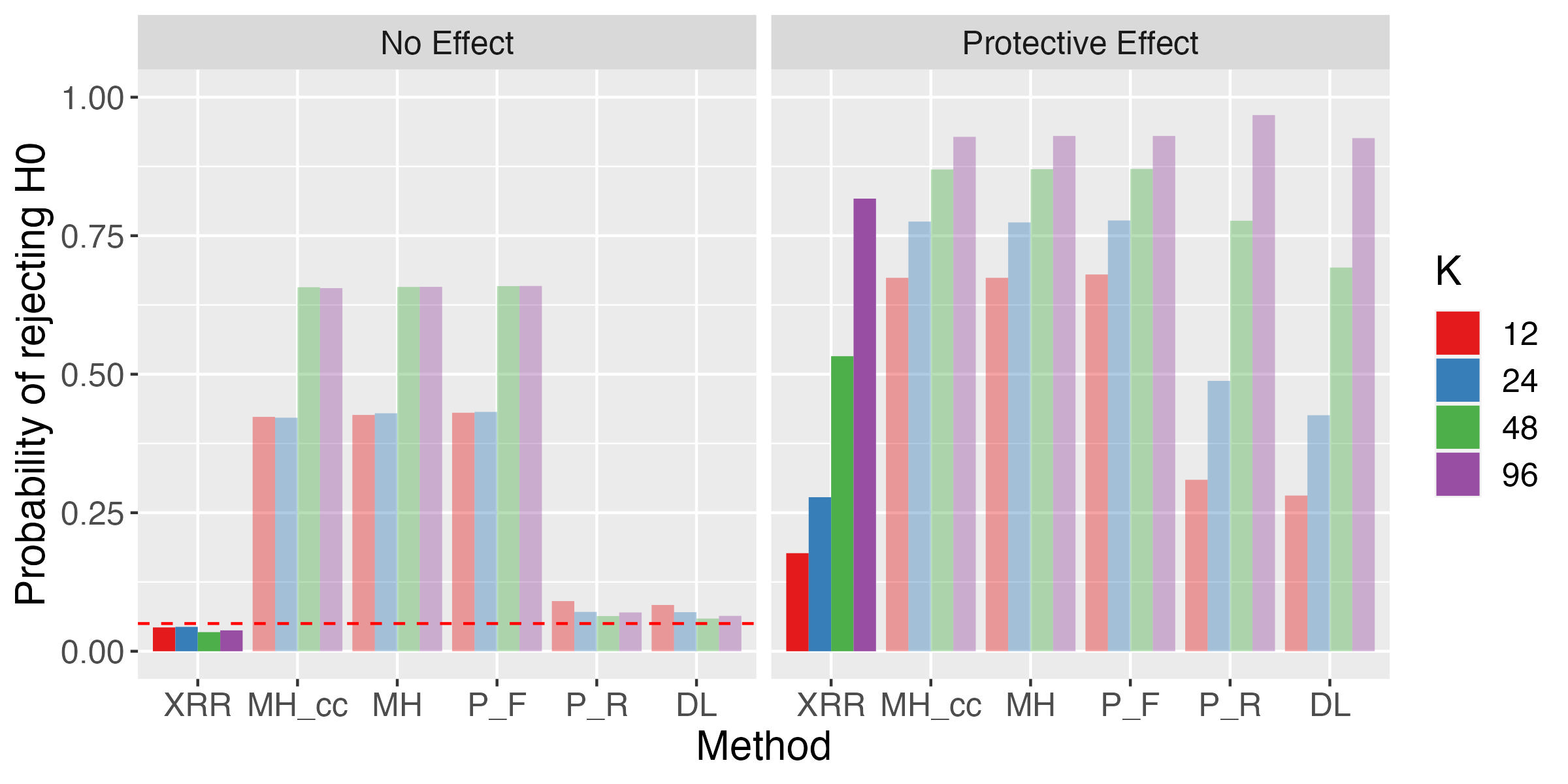}
\end{subfigure}
\begin{subfigure}[b]{\textwidth}
\centering
\caption{Setting 2: Moderate Heterogeneity}
\includegraphics[scale = 0.12]{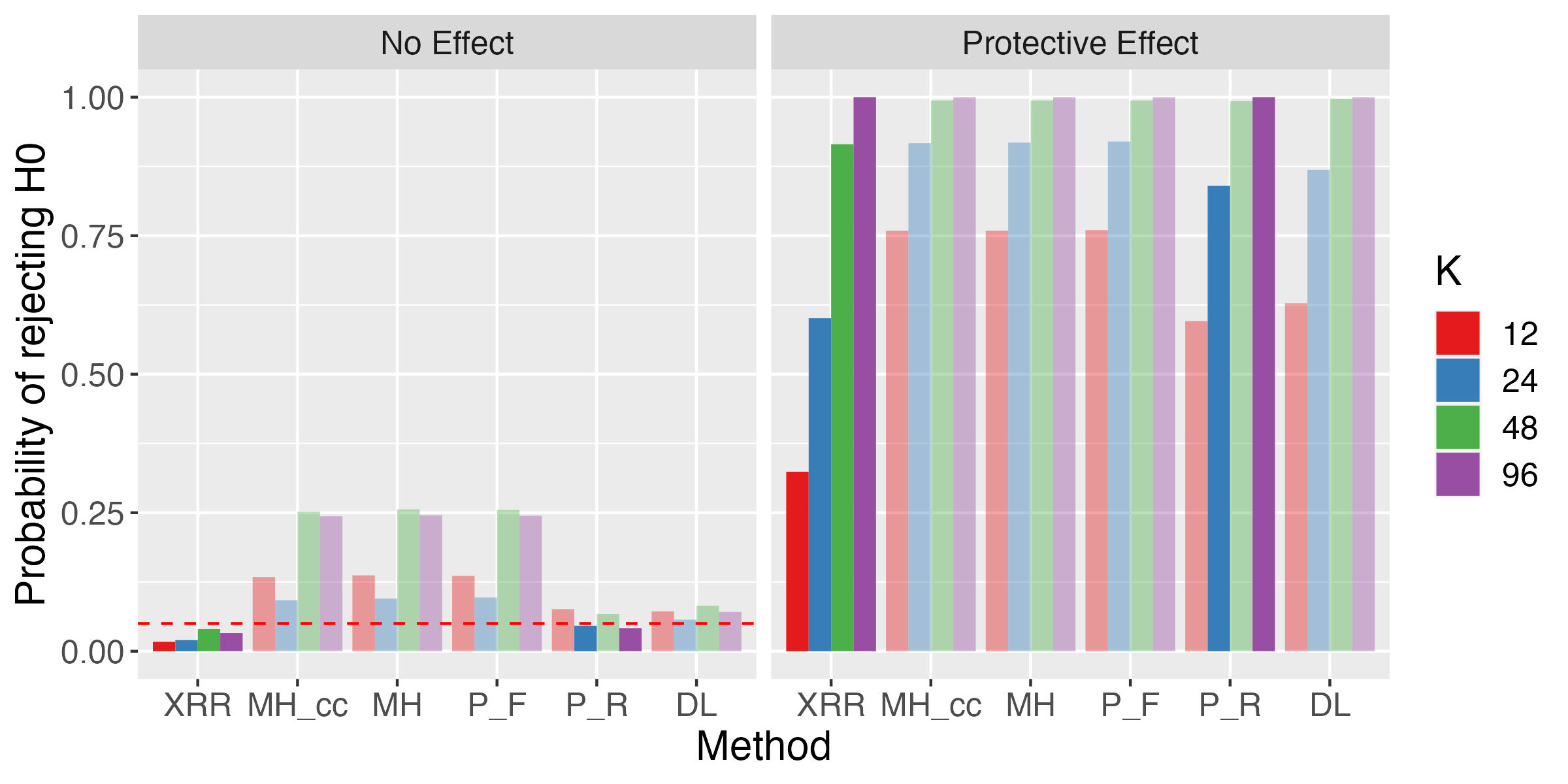}
\end{subfigure}
\begin{subfigure}[b]{\textwidth}
\centering
\caption{Setting 3: Low Heterogeneity}
\includegraphics[scale = 0.12]{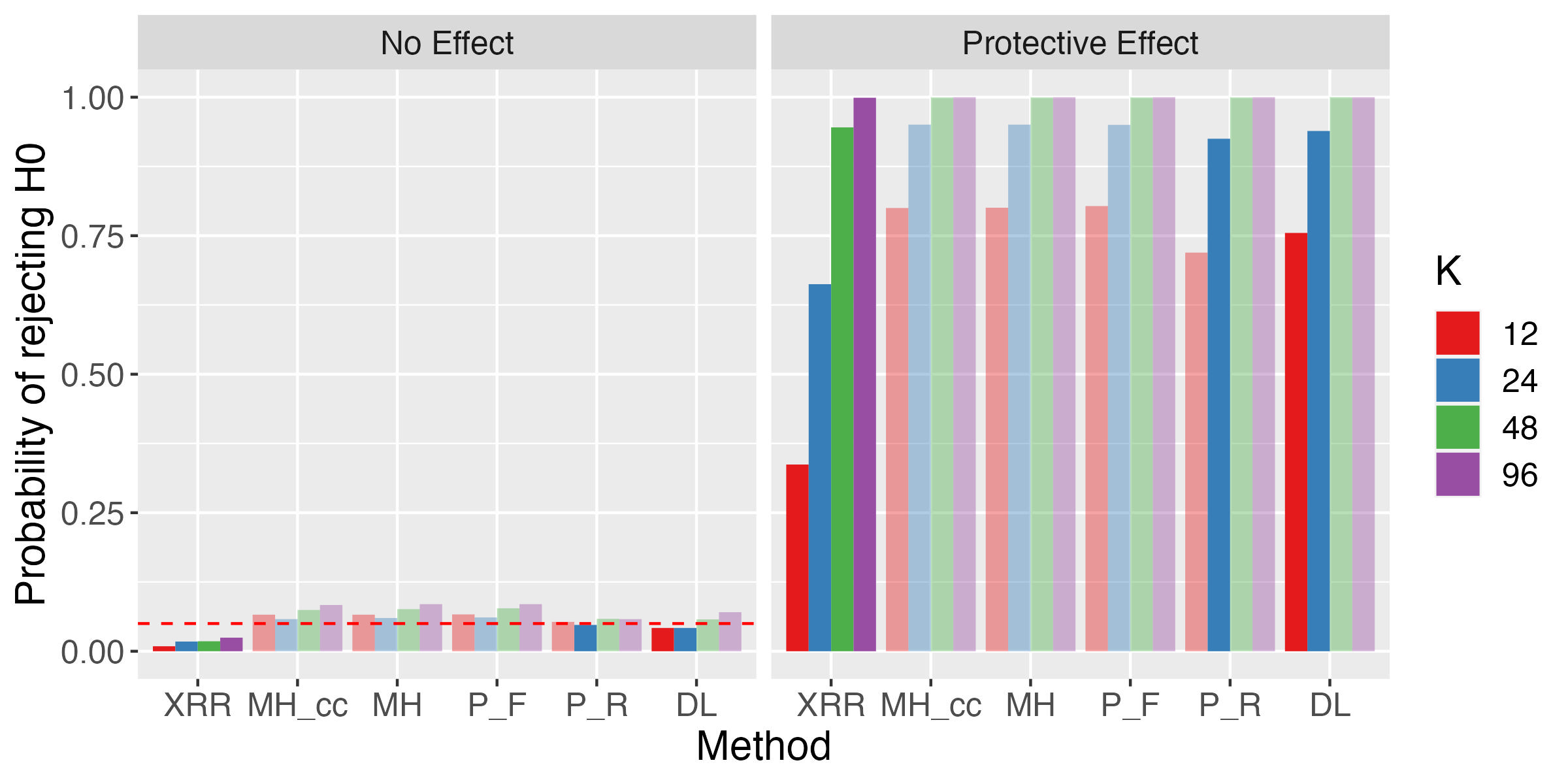}
\end{subfigure}
\caption{Type I error and power with $r_0 = 0.03$ for XRRmeta (XRR), Mantel-Haenszel with and without a 0.5 continuity correction (MH, MH-CC), the fixed and random effects Peto method (Peto-F, Peto-R), and the DerSimonian-Laird method with a 0.5 continuity correction (DL).  Methods that do not control the type I error are shown in a lighter shade.}\label{sim_result_supp}
\end{figure}

\newpage
\section{Face Mask Data}

\begin{table}[h!]
\centering
\scalebox{0.75}{\begin{tabular}{rrrrr}
  \hline
  \hline
    & \multicolumn{2}{c}{Face Mask} & \multicolumn{2}{c}{No Face Mask} \\
\hline
 Study ID & $N$ & Transmission Events & $N$ & Transmission Events\\ 
  \hline

1 &  16 &   3 &  15 &   4 \\ 
  2 & 123 &   8 & 354 &  43 \\ 
  3 &  98 &  11 & 115 &  61 \\ 
  4 & 202 &  46 &  55 &  31 \\ 
  5 &  24 &   3 &   4 &   2 \\ 
  6 &   7 &   0 &   2 &   1 \\ 
  7 &  31 &   0 &   6 &   3 \\ 
  8 &  43 &   8 &  72 &  17 \\ 
  9 &  61 &  17 &  18 &  14 \\ 
  10 &  42 &   8 &  25 &  14 \\ 
  11 &  23 &   3 &   9 &   5 \\ 
  12 & 278 &   0 & 215 &  10 \\ 
  13 &  51 &   0 & 203 &  13 \\ 
  14 & 1286 &   1 & 4036 & 119 \\ 
  15 & 116 &   6 & 101 &  12 \\ 
  16 &  62 &   2 &  10 &   2 \\ 
  17 &  26 &   3 &  60 &  33 \\ 
  18 &  27 &   6 &  71 &  39 \\ 
  19 & 218 &   0 & 230 &   6 \\ 
  20 & 444 &   1 & 308 &  16 \\ 
  21 &  42 &   0 &   6 &   0 \\ 
  22 &  24 &   0 &  10 &   0 \\ 
  23 &  60 &   0 &  45 &   0 \\ 
  24 &  13 &   0 &  19 &   0 \\ 
  25 &  64 &   0 &  13 &   0 \\ 
  26 &  61 &   0 &   1 &   0 \\ 
  27 &  89 &  12 &  98 &  25 \\ 
  28 & 146 &  25 & 229 &  69 \\ 
  29 &   9 &   0 & 154 &   7 \\ 
   \hline
   \hline
\end{tabular}}
\caption{Data for the face mask study. Shown are the study sizes (N) and the number of transmission events for the face mask and no face mask arms.}\label{tab: covid data}
\end{table}